\documentclass[%
reprint,
superscriptaddress,
amsmath,amssymb,
aps,
prapplied,
]{revtex4-2}

\usepackage{graphicx}
\usepackage{dcolumn}
\usepackage{soul}
\usepackage{bm}
\usepackage{physics}
\usepackage{verbatim} 
\usepackage[hidelinks]{hyperref}
\hypersetup{
    colorlinks=true,    
    allcolors=blue,    
    }

\begin{document}
	
    \preprint{APS/123-QED}
	
    \title{Optimal control of spin qudits subject to decoherence using amplitude-and-frequency-constrained pulses}	
	
    \author{Alonso Hern{\'{a}}ndez-Ant{\'{o}}n}
    \affiliation{Departamento de F{\'{\i}}sica de la Materia Condensada, Universidad de Zaragoza, 50009 Zaragoza, Spain}
    \affiliation{Instituto de Nanociencia y Materiales de Arag{\'{o}}n (INMA), CSIC-Universidad de Zaragoza, Pedro Cerbuna 12, 50009 Zaragoza, Spain}
    \affiliation{Department of Physics, ETH Zurich, Zurich CH-8093, Switzerland} \email{ahernandez@phys.ethz.ch}
	
    \author{Fernando Luis}
    \affiliation{Departamento de F{\'{\i}}sica de la Materia Condensada, Universidad de Zaragoza, 50009 Zaragoza, Spain}
    \affiliation{Instituto de Nanociencia y Materiales de Arag{\'{o}}n (INMA), CSIC-Universidad de Zaragoza, Pedro Cerbuna 12, 50009 Zaragoza, Spain}
	
    \author{Alberto Castro}
    \affiliation{Departamento de F{\'{\i}}sica Te\'orica, At\'omica y Optica, Universidad de Valladolid, 47011 Valladolid, Spain}
    \affiliation{Institute for Biocomputation and Physics of Complex Systems (BIFI) of the University of Zaragoza, 50018 Zaragoza, Spain}
	\email{alberto.castro.barrigon@gmail.com}
	\date{\today}
	
	\begin{abstract}
            Quantum optimal control theory (QOCT) can be used to design the shape of electromagnetic pulses that implement operations on quantum devices.
            By using non-trivially shaped waveforms, gates can be made significantly faster than those built by concatenating monochromatic pulses.
            Recently, we applied this idea to the control of molecular spin qudits modelled with Schrödinger's equation and showed it can speed up operations, helping mitigate the effects of decoherence [Phys. Rev. Appl. {\bf 17}, 064028 (2022)].
            However, short gate times require large optimal pulse amplitudes, which may not be experimentally accessible.
            Introducing bounds to the amplitudes then unavoidably leads to longer operation times, for which decoherence can no longer be neglected.
            Here, we study how to improve this procedure by applying QOCT on top of Lindblad's equation, to design control pulses accounting for decoherence already in the optimization process.
            We define the control signal in terms of generic parameters, which permits the introduction of bounds and constraints. This is convenient, as amplitude and frequency limitations are inherent to waveform generators.
            The pulses that we obtain consistently enhance operation fidelities compared to those achieved with the optimization based on Schrödinger's equation, demonstrating the flexibility and robustness of our method.        
            The improvement is larger the shorter the spin coherence time $T_{2}$.
	\end{abstract}

\maketitle
	
\section{\label{sec:intro}Introduction}

The main challenge in the path towards practical quantum computation is to keep noise
sufficiently low while scaling up the circuit volumes that can be executed. This can be 
tackled with the application of error correction codes
\cite{knill_theory_1997}. Most experimental demonstrations
\cite{postler_demonstration_2022, bluvstein_quantum_2022,
	abobeih_fault-tolerant_2022, takeda_quantum_2022,
	acharya_suppressing_2023} use a large number of physical qubits to
encode a small amount of logical information. Besides, they require
many nonlocal operations, a large number of control lines and complex
room-temperature electronics or lasers. A promising alternative makes
use of qudits, in which more than two energy levels of a physical unit
are used to encode logical information \cite{gottesman_encoding_2001,
	leuenberger_quantum_2001, brennen_criteria_2005}.
These could enable the integration of nontrivial operations within
single physical units \cite{lanyon_simplifying_2009,
	kiktenko_single_2015}, including error correction protocols
\cite{pirandola_minimal_2008,chiesa_molecular_2020}, which would entail a competitive
advantage. Qudits have been realized with
multiple physical systems, including photons
\cite{lapkiewicz_experimental_2011}, trapped ions
\cite{ringbauer_universal_2022}, impurity nuclear spins in
semiconductors \cite{asaad_coherent_2020} or superconducting circuits
\cite{neeley_emulation_2009}.

Gate operations in such multilevel systems are often designed with sequences of monochromatic resonant electromagnetic pulses, whose total duration becomes longer as the maximum allowed amplitudes get lower. These operations are naturally affected by errors, coming from two sources. One is leakage to undesired levels and the breakout of the rotating-wave approximation (RWA), and it can be alleviated by using low amplitudes, thus necessarily long durations. Another source is decoherence, and the way to mitigate it is exactly the opposite, i.e. using larger amplitudes to speed up the operation times with respect to the coherence time. One must find a compromise between these two, but depending on the particular system and the strength of the decoherence, this may not give good enough operation fidelities.

Quantum optimal control theory (QOCT)~\cite{brif_control_2010,
  glaser_training_2015, Koch2022} is a set of theoretical results and
algorithms for the study the following problem: given a
quantum system whose evolution can be controlled by some external
perturbations, find the optimal shape for those perturbations such
that a merit function of the system evolution is maximized.  Although
initially mostly used to generate state-to-state transitions, it
was later demonstrated~\cite{Palao2002, Palao2003} that it can also be used
with the goal of generating unitaries, i.e. quantum gates. Instead of sequences of monochromatic pulses, one can find
non-trivial, hopefully faster, temporal shapes for electromagnetic
pulses that produce the target gates.

Here, we build on this idea by developing a QOCT-based method for open quantum systems with the purpose of generating quantum gates.
Our approach and its results are illustrated with calculations performed for a specific, and quite natural, qudit realization: molecular nanomagnets with spin $S>1/2$ that are controlled with microwave magnetic pulses~\cite{gatteschi_quantum_2003, aromi_design_2012, atzori_second_2019, gaita-arino_molecular_2019, carretta_perspective_2021}. 

In a previous work~\cite{castro_optimal_2022}, we applied quantum optimal control theory (QOCT)~\cite{brif_control_2010, glaser_training_2015, Koch2022} to implement a target unitary operation $U_{\rm target}$, modelling the dynamics of the system with Schrödinger's equation (i.e. for closed systems), and we demonstrated a systematic improvement of fidelities in comparison to monochromatic sequences.
The particular QOCT method that we developed works on parametrized representations of the control functions, and allows us to easily implement bounds and constraints on their shape. The parameters can then be optimized with respect to a merit function by making use of an expression for the gradient derived with the adjoint method~\cite{Cao2003}.
We will refer to this method as QOCT-S hereafter. As it is natural, pulse duration and amplitude are inversely related for the optimal pulses: for a given amplitude (e.g. the largest experimentally attainable), a minimum duration must be allowed for the fidelity of the operation to be acceptable. While this duration can be significantly smaller than that required for monochromatic pulse sequences, the effect of decoherence may still be non-negligible. In other words, when the optimal pulses found with a model that ignores the environment are tested in the presence of decoherence, fidelities can get significantly reduced.

The aim of the present work is to develop a similar method to optimize the control pulses which takes into account decoherence.
Our goal is to study the interplay between control and incoherent errors, with the aim of bringing gate fidelities closer to the coherence limit.
For this purpose we generalize the methododology developed in~\cite{castro_optimal_2022}: in this case,
we apply QOCT on top of Lindblad's master equation. We will refer to this approach as QOCT-L.
As in the previous case, our parametrization allows us to set bounds on amplitude and frequency, which reflect the limitations that are inherent to experimental electronics.

Most of the methods based on QOCT, and the corresponding calculations
presented up to date have used techniques based on the manipulation of
the control functions in real time. By this we mean that the
parameters of the optimization are the values that the functions take
at each point in time during the process. In fact, this is not only
true in the quantum case, but it can also be said of optimal control
in general: Pontryagin's maximum principle~\cite{Pontryagin1962}, at
the core of the theory, is formulated in terms of the direct
optimization of the control functions, and not of any possible set of
parameters that may be used to specify those functions. While this
distinction may seem subtle, it has led to the development of a
majority of algorithms that do not parametrize the functions -- unless
we consider the values of the function in real time as the parameters.
Those methods have difficulties when one attempts to constrain the
control functions in some way (frequencies, amplitudes, etc.)~\cite{
[{Despite those difficulties, some procedures have been proposed to
add spectral constraints to real-space formulations of QOCT; 
see for example: }] [{ and also: }] Reich2014,*Lapert2009}.
Those constraints are needed if the calculations are to be useful for
experimental or technological purposes. For this reason, a (scarce)
number of methods based on generic or specific parametrizations -- such as this one -- have
been put forward over the last years~\cite{Castro2012,Machnes2018,Lucarelli2018,Sorensen2018}.

The manuscript is organized as follows. In Sec.~\ref{sec:model} we
present the model for a GdW${}_{30}$ molecular spin qudit with $d=8$ levels, arising from 
its $S=7/2$ spin \cite{jenkins_coherent_2017}, and 
describe the control of this system with sequences of monochromatic
pulses. In Sec.~\ref{sec:oct} we describe the QOCT methodology used in
this work, which takes into account decoherence. In
Sec.~\ref{sec:results} we show results of this QOCT-L optimization and
compare them to the ones obtained with the QOCT-S procedure,
and to monochromatic sequences. Finally,
Sec.~\ref{sec:conclusions_outlook} summarizes the conclusions.

\section{\label{sec:model}The GdW${}_{30}$ molecular spin qudit: control with monochromatic resonant pulses}

\begin{figure}
	\centering
	\includegraphics[width = \linewidth]{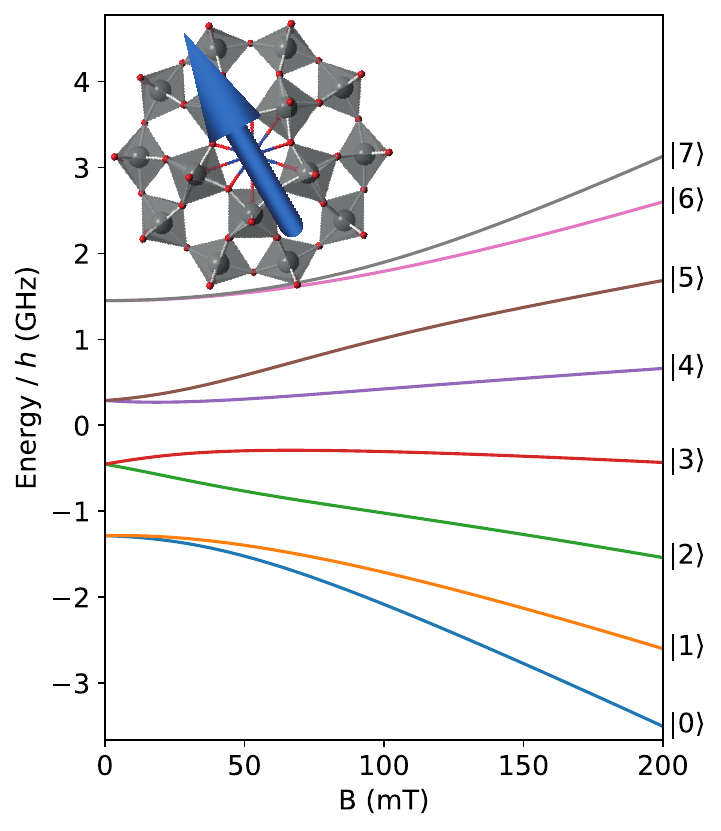}
	\caption{Scheme of the $8$ spin energy levels of the GdW$_{30}$ complex, whose molecular structure is shown in the inset, for a DC magnetic field applied along its medium anisotropy axis $x$.}
	\label{fig:molecule_levels}
\end{figure}

We consider a model molecular spin qudit, the GdW${}_{30}$
complex~\cite{jenkins_coherent_2017}, whose core is schematically shown in the inset of Fig.~\ref{fig:molecule_levels}. This core consists of a single 
Gd$^{3+}$ ion with a $4f^7$ configuration, whose ground manifold 
has $L = 0$ and $S = 7/2$. This spin manifold gives the basis for  
encoding a $d = 8$ \textit{qudit}. The interaction with the 
polyoxometalate moieties surrounding the Gd$^{3+}$ ion gives rise 
to a weak, yet finite magnetic anisotropy \cite{martinez-perez_gd-based_2012}. Under the effect of a DC 
magnetic field $\vec{B}$, the spin Hamiltonian of this molecule 
can be well approximated by an
orthorhombic zero field splitting plus a Zeeman contribution
\cite{White2007}:
\begin{equation}
	{\cal H}_0=D\left[S_z^2-\frac{1}{3}S(S+1)\right]+E\left(S_x^2-S_y^2\right)-g\mu_\text{B} \vec{S}\cdot\vec{B},
	\label{eq:Hamiltonian}
\end{equation}
where $S_x$, $S_y$ and $S_z$ are spin operators, and
$D=1281~\text{MHz}$ and $E=294~\text{MHz}$
are magnetic 
anisotropy constants. After diagonalization, the Hamiltonian may be written as
\begin{equation}
	{\cal H}_0(B) = \sum_{n = 0}^{d-1} E_n(B) \ket{n}\bra{n}.
\end{equation}
\noindent where $E_{n}(B)$, with $n=0,\ldots,7$, are the energy eigenvalues, shown in Fig.~\ref{fig:molecule_levels} as a function 
of magnetic field, and $\vert n \rangle$ their corresponding eigenstates. In all the calculations that follow, the 
magnetic field is oriented along the medium magnetic axis $x$, 
that is, $\vec{B} = B \hat{x}$. This orientation minimizes the 
dispersion between the frequencies of resonant transitions linking 
adjacent levels, 
$\vert n \rangle \rightarrow \vert n \pm 1 \rangle$, which are the 
only ones allowed for 
sufficiently high $B$ \cite{jenkins_coherent_2017}. Still, as can be seen in Fig.~\ref{fig:molecule_levels}, all these frequencies 
remain different on account of the magnetic anisotropy, which allows for full addressability.

{
In this work, we assume that the interaction of the spin qudit with the environment, chiefly the nuclear spins in neighboring atoms, can 
be described within a  Markovian approximation~\cite{lindblad_generators_1976,gorini_completely_1976}. This approximation is quite common in theoretical studies of spin decoherence  
in molecular systems~\cite{chiesa_molecular_2020, chiesa_theoretical_2022, chiesa_molecular_2024, jankovic_noisy_2024}. It provides a suitable method 
to introduce the effects of dephasing and dissipation on the spin dynamics 
while keeping the equations, and in our case the optimization problem, 
tractable. Besides, it seems 
justified for this specific molecule: as shown in \cite{jenkins_coherent_2017} the Rabi oscillation amplitudes corresponding 
to each of the $d-1$ allowed resonant spin transitions decay quasi-
exponentially with time. The spin dynamics can be then described by a 
Lindblad equation:}
\begin{equation}
	\label{eq:lindblad}
	\dot{\rho} = \mathcal{L}\rho = -i \left[{\cal H}_0, \rho\right] + \sum_\alpha \gamma_\alpha
	\left(
	L_\alpha\rho L_\alpha^\dagger - \frac{1}{2}\lbrace L_\alpha^\dagger L_\alpha, \rho\rbrace
	\right)
\end{equation}
The dissipative term is therefore defined by a set of Lindblad 
operators
$L_\alpha$, each of them associated to a decoherence rate
$\gamma_\alpha$. For the spin qudits that we are considering, pure dephasing is the dominant decoherence
effect \cite{petiziol_counteracting_2021}. Therefore, we have limited our numerical 
experiments to the case in which the only nonzero Lindblad 
operators are $L_\phi^n = \ket{n}\bra{n}$, all of them associated
to the same dephasing constant $\gamma_\phi = 1/T_{2}$. The latter 
condition agrees with the results of spin-echo experiments performed on crystals of GdW$_{30}$ diluted in diamagnetic YW$_{30}$, which show that $T_{2}$ is nearly the same for all resonant transitions \cite{jenkins_coherent_2017}.

The model described by Eqs.(\ref{eq:Hamiltonian}) and 
(\ref{eq:lindblad}) includes no external driving. Let us hereafter 
assume that we add some extra time-dependent term to ${\cal H}_0$, 
so that the total spin Hamiltonian becomes:
\begin{equation}
	{\cal H}(t) = {\cal H}_0 + f(t)V,
	\label{eq:perturbed_hamiltonian}
\end{equation}
where $V$ describes the interaction mechanism between the control
electronics and the molecule, and $f(t)$ is the temporal shape
of the drive signal. In the following, we use a time-dependent
magnetic field oriented along the easy anisotropy axis $y$ (thus perpendicular to the static one), i.e. $V = -g\mu_\text{B} S_y$.

The external control drive is designed through the function $f(t)$
of Eq. (\ref{eq:perturbed_hamiltonian}). Different
approaches to the control of the system boil down to choosing
different functional shapes of $f(t)$. Coherent control of quantum
devices is typically implemented by applying sequences of
monochromatic pulses, each one addressing the transition between 
two levels. The result
of these pulses is most conveniently pictured as rotations in the
Bloch sphere defined in each two-level subspace. A generic rotation
(which is a unitary operation) can be parameterized as:
\begin{equation}
	R^{(jk)}_{\vec{n}}(\theta) = \exp\left(-i\frac{\theta}{2}\vec{n}\vec{\sigma}^{(jk)} \right)
\end{equation}
Here, $\vec{\sigma}^{(jk)}$ is the vector of Pauli matrices within 
the subspace spanned by the two levels $(jk)$ -- in the basis of
eigenstates of the control-free Hamiltonian ${\cal H}_0$. The unit 
vector $\vec{n}$ determines the rotation axis and $\theta$ is the 
rotation angle.

In the interaction representation~\footnote{In this work, we will
assume that the purpose is to realize a given gate in the 
interaction representation.}, this unitary operation can be 
\emph{approximated} via the application of a pulse in the form:
\begin{equation}
f(t) = A \Pi_0^{t_f}(t) \cos(\omega_{jk} t + \phi),
\end{equation}
where
$\Pi_{t_a}^{t_b}(t)$ is a square envelope in the interval $(t_a, 
t_b)$ and $\omega_{jk} = \vert E_{k}-E_{j}\vert / \hbar$. The 
amplitude $A$ must be chosen in combination with
$t_f$, such that:
\begin{equation}
\label{eq:lambdaT1}
A t_f = \frac{\theta}{\vert \langle j \vert V \vert k \rangle\vert}
\end{equation}
It is clear from this equation that, the larger the amplitude
$A$, the shorter the duration $t_f$ that is required to complete the rotation. Finally, in order to have a 
rotation around the $\vec{n}$ direction, the phase $\phi$ must be chosen such that:
\begin{equation}
\vec{n} = (\cos(\phi - \arg \langle j \vert V \vert k \rangle), \sin(\phi - \arg \langle j \vert V \vert k \rangle), 0 )
\end{equation}
Note that $n_z = 0$:
rotations around the $z$ axis cannot be directly realized in this
way, but they can be built through the composition of three 
rotations around the $x$ and $y$ axes:
\begin{equation}
\label{eq:zdecomp}
R_Z(\theta) = R_X(\pi/2)R_Y(\theta)R_X(-\pi/2)
\end{equation}

In general, any unitary operation can be decomposed into a 
sequence of rotations: for any target unitary $U_{\rm target}$, one
can always find a set of rotations such that:
\begin{equation}
U_{\rm target} = R^{(j_rk_r)}_{\vec{n}_r}(\theta_r) \dots R^{(j_2k_2)}_{\vec{n}_2}(\theta_2) 
R^{(j_1k_1)}_{\vec{n}_1}(\theta_1)\,,
\end{equation}
modulo a global unimportant phase factor.
In order to implement this, one can then define a control waveform consisting of a sequence of the monochromatic single pulses defined above.
If the amplitude $A$ for all the pulses in the sequence is 
the same, fixed by experimental limitations, then it must be 
related to the total duration of the sequence $T$ by the relation:
\begin{equation}
\label{eq:lambdaT}
A T = \sum_{i=1}^r \frac{\theta_i}{\vert\langle j_i \vert V \vert k_i\rangle\vert}\,.
\end{equation}
It is clear how one may then concatenate multiple 
monochromatic single pulses that implement any given unitary
operation~\footnote{
An implicit assumption in Eq.~(\ref{eq:lambdaT}) (and in the discussion around it) is the possibility
of discontinouisly switching from one monochromatic pulse to another. In practice, these pulses
require some switch-on and off phases, that make the real implementations slower. Expressions
(\ref{eq:lambdaT1}) and (\ref{eq:lambdaT}) can thus be considered lower bounds for the durations}. 
This can be done by fixing either the total duration $T$ or the amplitude $A$.

Note, however, that even in the absence of decoherence the 
resulting operation will not exactly match $U_{\rm target}$:
the  previous discussion relies on the RWA and on the absence of leakage to other
levels during each two-level operation. Reducing this error 
requires that all 
transition frequencies are sufficiently different to each other, 
and that the amplitude $A$ is low enough to make the RWA hold: this 
approximation becomes exact only in the limit of vanishing 
amplitude -- and, therefore, infinitely long durations. The 
constructed operation will
then differ from the target unitary, and the difference will be larger 
if shorter durations are demanded. 

This sets a limit on the 
operation speeds that can be implemented by concatenating monochromatic pulses: one cannot 
simply increase the amplitude $A$ (even if it were 
experimentally possible) in order to reduce the operation times, 
since the error due to the approximate nature of
the expressions above would then increase. To minimize this error, 
one should use low amplitudes and long operation times. But then, the system becomes more prone to the errors caused by 
decoherence. It
is clear how the two kinds of error cannot be simultaneously
minimized. One must find a compromise amplitude, which
may or may not lead to an acceptable fidelity in the operation
depending on the level of noise, system characteristics, etc.

The above discussion justifies searching for an alternative to
decomposing the unitary into rotations implemented by monochromatic
pulse sequences: one may then think of creating the target unitary with a
complex multi-frequency pulse tailored with QOCT.  In
\cite{castro_optimal_2022}, we tried this route using an optimization
procedure that did not account for decoherence. In the next sections,
we describe how this methodology can be improved by including it.

\section{\label{sec:oct}Quantum Optimal Control Theory for Open Systems}

In this section, we describe the mathematical procedure that applies
QOCT to a system coupled to a dephasing bath in order to realize a
given target gate $U_{\rm target}$. The application of QOCT to open
quantum system was reviewed by Koch~\cite{Koch_2016}.  See for example
\cite{Schulte-Herbruggen_2011,goerz_optimal_2014} for early
applications of this method for the purpose of creating unitary
operations in the presence of decoherence. The scheme outlined here,
based on Eqs.~(\ref{eq:gradientfinal}) and (\ref{eq:costate_lind_multitarget}), 
permits to use general parametrizations of the control functions -- in contrast
to the more common real-time representation used by many approaches. This facilitates
the enforcement of experimental constraints, as we will show below.

First, we define a 
functional form for the control pulse $f = f(u, t)$,
determined by a set of control parameters $\{u_k\}\equiv u$. The 
time-dependent spin Hamiltonian (\ref{eq:perturbed_hamiltonian}) 
is then rewritten as:
\begin{equation}
	{\cal H}(u, t) = {\cal H}_0 + f(u, t)V\,.
\end{equation}
The parameterized form of the control functions that we use is:
\begin{equation}
	f(u, t) = \Phi(\tilde{f}(u, t))\,,
	\label{eq:fut}
\end{equation}
where $\tilde{f}$ is the Fourier expansion
\begin{equation}
\label{eq:Fourierexpansion}
	\tilde{f}(u, t) = \sum_{m=1}^{M} 
	\left( u_{2m}\cos(\omega_m t) + u_{2m-1}\sin(\omega_m t)\right)\,.
\end{equation}
Here $\omega_m = \frac{2\pi}{T}m$, and $M$ sets a cutoff frequency $\omega_\text{max}$ that can be chosen according
to the sampling rate of the waveform generator. In order to bound the amplitude of the signal, this expansion 
is modulated by a function $\Phi$ defined in such a way that:
\begin{equation}
	\vert f(u, t)\vert \le \kappa\quad\textrm{ at any $t$.}
\end{equation}
The design of $\Phi$ is explained in Appendix~\ref{appendix:phi}. The goal is to fix both
bandwidth and amplitude limits for the control field.

Note that most applications of QOCT do not employ 
parameterized forms of the control functions, but rather work with 
the full \emph{real-time} representation of the function (one may
say that the parameters are directly the values of the function at the
discretized time grid used for the calculations). Although there are
also ways to constrain frequencies or amplitudes when using
such real-time formulations, using parameterized
forms like the one used here is more effective.

In what follows, we describe how we have employed QOCT to find the 
optimal shape $f(u, t)$ for the control of a spin qudit subject to 
decoherence. Given a choice of parameters $u$, the system 
departs from $\rho_0$ and evolves into $\rho(u, T)$, following 
Lindblad's equation:
\begin{equation}
	\dot{\rho} = \mathcal{L}(u, t)\rho\,,
\end{equation}
where, as compared to Eq. (\ref{eq:lindblad}), the dependency of 
the Lindbladian on the control parameters $u$ and on time has now 
been made explicit.

Any formulation of QOCT must define a \emph{merit} function for 
the system evolution. Let us consider first a simpler problem than 
the one that we really need to solve: say that we 
want to find a pulse shape that
drives the system from an initial pure state $\rho_0 =
\vert\psi_0\rangle\langle\psi_0\vert$ to an also pure final state $\rho_{\rm
	target}=\vert\psi_{\rm target}\rangle\langle\psi_{\rm target}\vert$.
In this case we can use a simple merit function:
\begin{equation}
	\label{eq:targetfunction}
	F(\rho) = \text{Tr}\left[ \rho \cdot \rho_{\rm target}\right]\,,
\end{equation}
This function has a maximum of one at $\rho = \rho_{\rm target}$. 
The problem then boils down to the maximization of the following 
function:
\begin{equation}\label{eq:fidelity}
	G(u) = F(\rho(u, T))\,,
\end{equation}
where $T$ is the total propagation
time, over all sets of parameters $u$. Despite some caveats related to 
the state purity~\cite{goerz_optimal_2014}, this
function provides a good measure of the fidelity
between the propagated state and the target. In order to compute 
$G(u)$, one must perform the numerical
propagation $\rho_0 \to \rho(u,T)$. One can then use any
of the many available optimization algorithms that do not
require the gradient of the function (\emph{gradient-free}). 
However, and specially since the number of parameters $u$ can be 
large, the search is more efficient if one uses a \emph{gradient-based} algorithm. Using QOCT \cite{pontryagin_mathematical_1962, castro_qocttools_2024},
one may then derive the following expression for the gradient:
\begin{equation}
	\label{eq:gradlind}
	\frac{\partial G}{\partial u_m}(u) = 2\int_0^T\!\!\!{\rm d}t\;  \frac{\partial f}{\partial u_m}(u, t) 
	{\rm Tr}\left(
	\lambda^\dagger(u, t)\big[\hat{V},\rho(u, t)\big]\right)
\end{equation}
In this equation, the \emph{costate} or \emph{adjoint state} $\lambda$ is defined by:
\begin{equation}
	\left\{\begin{array}{l}
		\dot{\lambda}(u, t)= -\mathcal{L}^\dagger(u,t)\lambda(u, t)\,,\\
		\lambda(u, T)=\frac{1}{2}\rho_{\rm target}\,.
	\end{array}\right.\label{eq:costate_lind}
\end{equation}
Numerically, the computation of the gradient using this expression
requires (1) solving the system equation of motion (Lindblad's
equation) for the state $\rho(u, t)$; (2) solving the previous
equation of motion for the costate $\lambda(u, t)$; and (3) 
computing the integral (\ref{eq:gradlind}).

However, the problem that we want to solve is more involved than
merely finding a pulse that executes the operation 
$\rho_0 \rightarrow \rho_{\rm target}$ for a certain state 
$\rho_0$. What we actually want to do is
to implement a unitary operation $U_{\rm target}$. The pulse 
should then map $\rho \rightarrow U_{\rm target} \rho U_{\rm 
	target}^\dagger$ for any possible initial state $\rho$. For this purpose, 
a more complex target function is needed. We take 
a set of states $\boldsymbol{\rho}^0 = \{\rho^0_k\}_{k=1}^N$ (ideally it should be a full basis), and 
define a target function that averages over the effect of the 
pulse on all those states:
\begin{equation}
	\label{eq:multitarget_fidelity_new}
	F(\boldsymbol{\rho}, u) = \frac{1}{\sum_k^N \Tr\left[(\rho_k^0)^2\right]}\sum_k^N \Tr\left[\rho_k(u, T)\rho_k^\text{target}\right] + P(u)\,,
\end{equation}
where
\begin{equation}
	\rho_k^\text{target} = U_\text{target}\rho^0_k U_\text{target}^\dagger\,.
\end{equation}
Note that we also allow for the presence of a {\it penalty} function $P(u)$: the definition
and purpose of this penalty are explained in Appendix~\ref{appendix:phi}.

Function $G$ is once again defined as
\begin{equation}
\label{eq:multitarget_fidelity}
G(u) = F(\boldsymbol{\rho}(u, T), u)\,.
\end{equation}
Except for the penalty, this is just an average over fidelities like the one defined in
(\ref{eq:fidelity}), the gradient equation is trivially extended from Eq.~(\ref{eq:gradlind}):
\begin{align}
\label{eq:gradientfinal}
	\frac{\partial G}{\partial u_m}(u) =& 2\sum_{k=1}^N 
	\int_0^T\!\!{\rm d}t\;  \frac{\partial f}{\partial u_m}(u, t) 
	{\rm Tr}\left(
	\lambda_k^\dagger(u, t)\big[\hat{V},\rho_k(u, t)\big]\right)
\\
        & + \frac{\partial P}{\partial u_m}(u)\,,
\end{align}
where
\begin{equation}
	\left\{\begin{array}{l}
		\dot{\lambda}_k(u, t)= -\mathcal{L}^\dagger(u,t)\lambda_k(u, t)\,,\\
		\lambda_k(u, T)=\frac{1}{2}\rho_k^{\rm target}\,.
	\end{array}\right.\label{eq:costate_lind_multitarget}
\end{equation}

The computation of this gradient is, however, more costly than
Eq.~(\ref{eq:gradlind}), as it requires $2N$ propagations: $N$ for 
the states $\rho_k(u, t)$, and $N$ for the corresponding costates
$\lambda_k(u, t)$. In principle $N=d^2$, the Liouville space
dimension. Fortunately, as shown in \cite{goerz_optimal_2014}, it 
is sufficient to define the fidelity using a minimum of three 
states. Here, we have adopted an alternative definition suggested in the same work, which is given in terms of 
$d+1$ states. These are the $d$ diagonal states $\{\rho_k = \ket{k}\bra{k}\}_{k = 0}^{d-1}$, which account for the proper mapping of populations, plus an additional
state $\rho_{d}$ defined as $(\rho_d)_{ij} = 1/d$, that permits to properly track the phases. This choice 
substantially reduces the computational cost of the optimization.

The method described above has been implemented in the {\tt 
	qocttools} code~\cite{castro_qocttools_2024}.

\section{\label{sec:results}Results}

\begin{figure*}
	\centering
	\includegraphics[width=\linewidth]{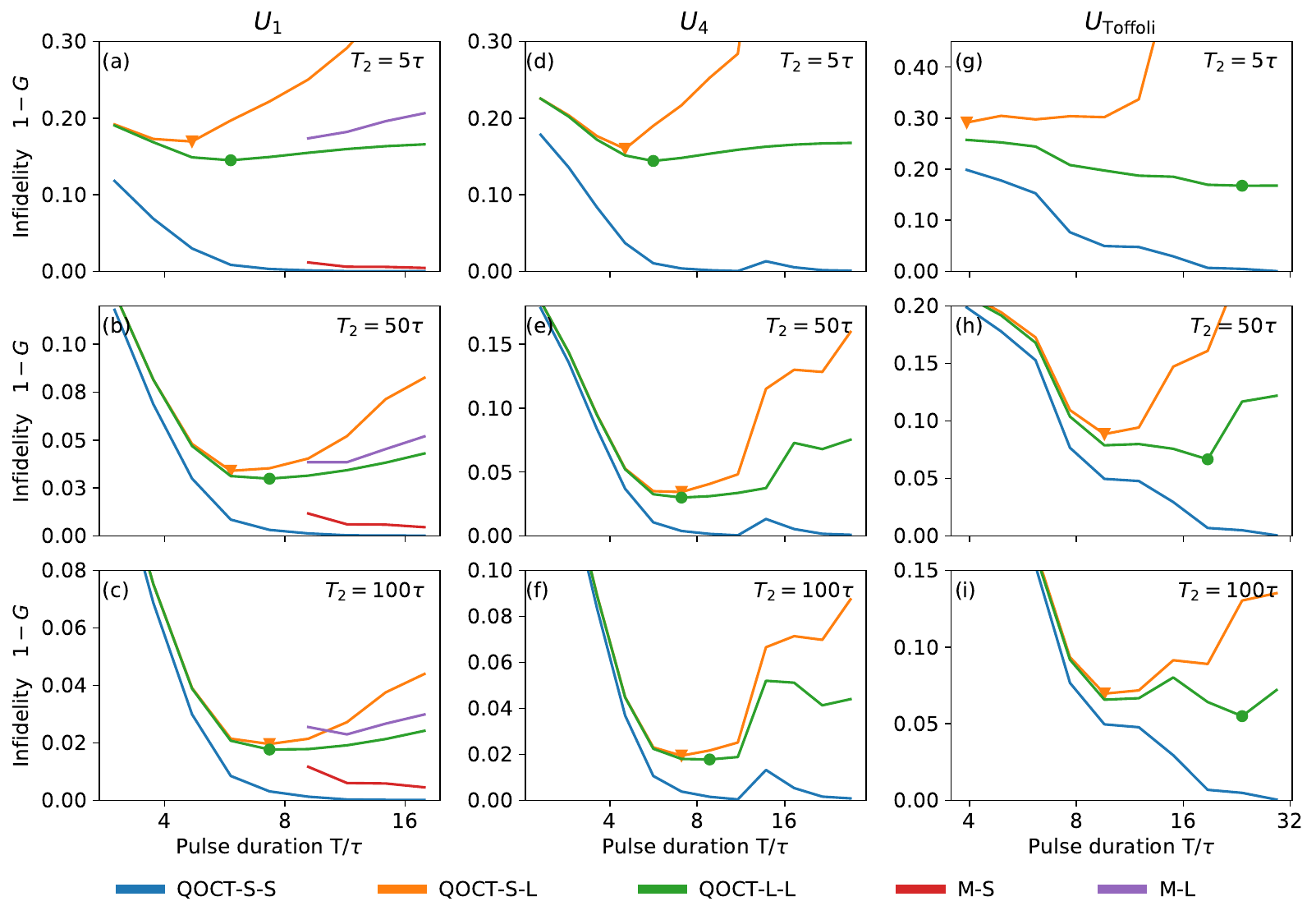}
	\caption{Infidelity of gate operations as a function of the total waveform length, for gates (a-c) $U_1$, (d-f) $U_4$ and (g-i) $U_\text{Toffoli}$. From top to bottom, the simulations use dephasing times of $T_2 = 5\tau, ~50\tau ~\text{and}~200\tau$ (inset). The results of monochromatic sequences start at longer durations because the amplitude constrain sets a minimum pulse length. The curves that are not displayed have larger durations or infidelities and are of less interest. The orange and green markers indicate the numerical minima of the QOCT-S-L and QOCT-L-L curves, respectively, indicating the best performance achieved with both methods in every case.}
	\label{fig:iF_T}
\end{figure*}

\begin{figure}
	\centering
	\includegraphics[width = \linewidth]{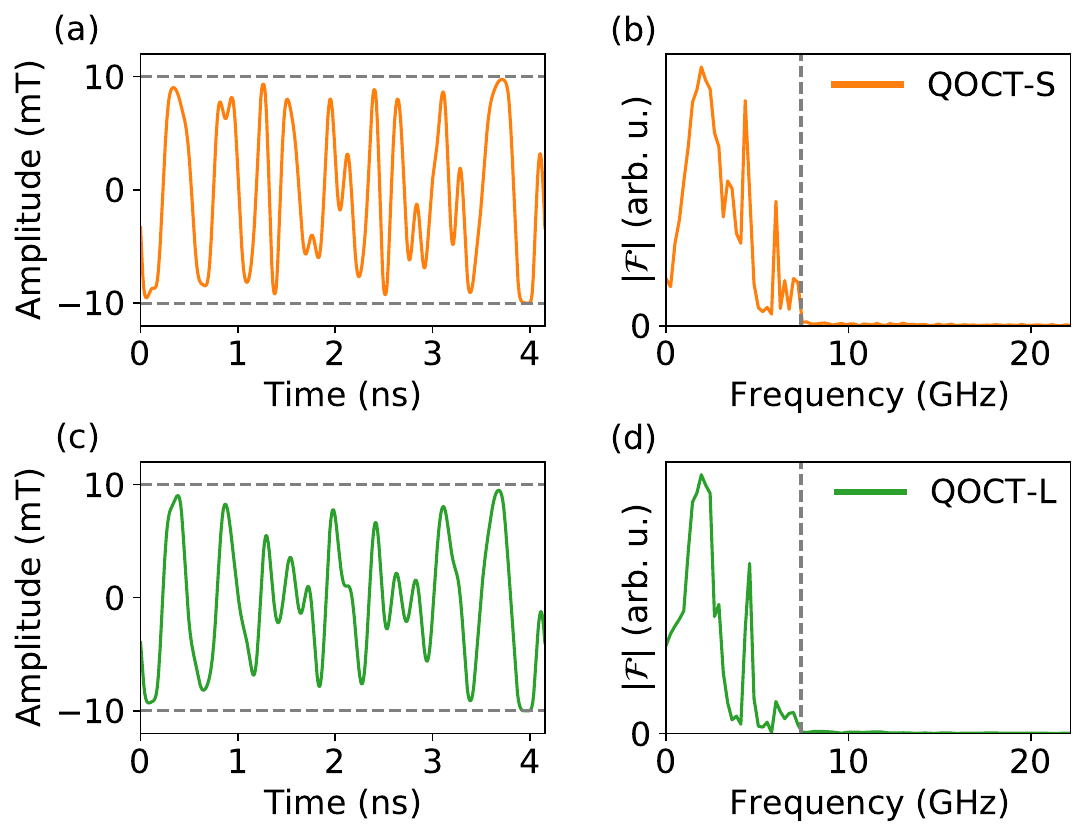}
	\caption{
    Pulses and their Fourier transforms obtained for the Toffoli gate with $T_2 = 50\tau$ and $T\approx 7.68\tau$ using methods (a, b) QOCT-S, (c, d) QOCT-L. The grey dashed lines indicate the amplitude and frequency bounds set for the optimization.
    }
	\label{fig:pulses}
\end{figure}

We apply the method described in the previous section to 
find pulses that implement, in the GdW$_{30}$ system, a given 
target unitary, in the presence of dephasing noise.
We set a static magnetic field of $B = 150~\text{mT}$, and set the following constraints
for the pulses:
a maximum frequency of $\omega_\text{max} = 4\omega_{67}$, 
where $\omega_{67}$ is the transition frequency between the levels 6 and 7,
and a maximum amplitude $A_{\rm max} = 10 \text{ mT}$.
We encode three qubits in the 8 energy levels of the spin
($\ket{0}\equiv\ket{000},~\ket{1}\equiv\ket{001},\dots,~\ket{7}\equiv\ket{111}$) and
choose the Toffoli gate as the target unitary, using qubits 1 and 2 as control and
qubit 3 as target. In order to realize this operation using a sequence of
monochromatic pulses, one would need to decompose it as a product of $8$
rotations: one $\pi$ rotation between levels
$\ket{6} = \ket{110}$ and $\ket{7}=\ket{111}$, plus phase gates ($R_Z$) on all
pairs of two adjacent levels:
\begin{align}\label{eq:toffolidecomp}\nonumber
	U_{\rm Toffoli} = e^{i\frac{\pi}{8}} &
	R_{Z}^{(01)}(\frac{1}{4}\pi)
	R_{Z}^{(12)}(\frac{1}{2}\pi)
	R_{Z}^{(23)}(\frac{3}{4}\pi)
	R_{Z}^{(34)}(\pi)
	\\
	&
	R_{Z}^{(45)}(\frac{5}{4}\pi)
	R_{Z}^{(56)}(\frac{3}{2}\pi)
	R_{Z}^{(67)}(\frac{3}{4}\pi)
	R_{X}^{(67)}(\pi)
\end{align}
Notice, however, that the last seven 
rotations around the $Z$ axis have to be implemented as a product 
of three $X$ and $Y$ rotations, as discussed above in
Eq.~(\ref{eq:zdecomp}). Therefore, the complete sequence consists of 
$22$ rotations, i.e. $22$ monochromatic pulses, even for a gate this
simple. This makes the total duration very large,
and lets decoherence come into play. Below, we will compare the performance of
this method based on the decomposition of the unitary into rotations and the use of monochromatic pulses, with the methods based on optimized multifrequency pulses,  QOCT-S and QOCT-L. In addition to using
the ``full'' Toffoli gate, and for the sake of
observing how the use of QOCT becomes more and more useful as the 
complexity of a gate (in terms of number of necessary 
rotations) grows, we have also considered shorter sequences, using the 
first rotations that enter the above definition of the Toffoli 
gate:
\begin{align}
	U_1 &= 
	R_{Z}^{(67)}(\frac{3}{4}\pi)
	\\
	U_4 &=
	R_{Z}^{(67)}(\frac{3}{4}\pi)
	R_{X}^{(67)}(\pi)
\end{align}
After decomposing each $Z$ rotation into three $X$ and $Y$
rotations, $U_1$, $U_4$ require $1$ and $4$ monochromatic pulses, respectively.

We have then followed the following computational protocol:
\begin{enumerate}
	
	\item For a series of predefined times $T$, we implement the target
	unitary as a sequence of monochromatic
	pulses, and compute the resulting fidelity using the merit
	function defined by Eq.~(\ref{eq:multitarget_fidelity}), \emph{in the
	absence of decoherence}. Even then, the fidelity is not one, as
	there is an error due to leakage and the RWA. This error should
	decrease with lower amplitudes, and thus larger $T$ according to
	Eq.~(\ref{eq:lambdaT}).	From this equation we also deduce that
	there is a minimum duration time, given by:
	\begin{equation}
                \label{eq:tmin}
		T_{\rm min} = \frac{1}{A_{\rm max}}
		\sum_{i=1}^r \frac{\theta_i}{\vert\langle j_i \vert V \vert k_i\rangle\vert}\,.
	\end{equation}
	We label these results obtained with monochromatic
	sequences and Schr{\"{o}}dinger's equation as ``M-S''.
	
	\item We then compute the corresponding operation fidelities
	obtained by those monochromatic sequences \emph{in the presence of
	decoherence}, using Lindblad's equation.  We label these
	results as ``M-L''.
	
	\item Next, we apply QOCT to Schr{\"{o}}dinger's equation to obtain an optimized pulse with the parameterization explained above, with the amplitude and frequency bounds $A_{\rm max}$ and $\omega_\text{max}$. We label these results as ``QOCT-S-S''.
	
	\item We then test the performance of the pulses optimized in the previous step, this time in the presence of decoherence, by solving the dynamics given by Lindblad's equation for those control waveforms. We label these results as ``QOCT-S-L''.
	
	\item Finally, we apply QOCT to Lindblad's equation as described above, to obtain optimized pulses with the same durations and amplitude and frequency bounds. We label these results as ``QOCT-L-L''.
	
\end{enumerate}

We use the Larmor period of the 6-7 transition $\tau =
\frac{2\pi}{\omega_{67}}$ as the unit of time. We consider, for each
target unitary, several values of $T_2$ ranging from $5\tau$ to
$200\tau$.  Fig.~\ref{fig:iF_T} displays some of the results obtained for
the gates $U_1$, $U_4$ and $U_\text{Toffoli}$ gates, where we quantify
the infidelity of the operations, $1-G$, using $G$ given by
Eq.~(\ref{eq:multitarget_fidelity}) as a measure of the fidelity.

Let us focus first on the simplest gate $U_1$, consisting of only one
rotation (left column). The red line (M-S) displays infidelities
obtained with the monochromatic sequences in the absence of
decoherence. Those lines start at a duration given by $T_{\rm min}$ in
Eq.~(\ref{eq:tmin}). These infidelities are not zero, but decrease as the
total duration times $T$ grow. In the presence of dephasing (M-L,
purple line), however, the error grows as $T$ becomes longer.

Using the QOCT-S-S method, one obtains the curves shown in blue. In this case the infidelities decay to zero with increasing $T$ faster than those obtained with monochromatic sequences. This proves how
the multi-frequency waveforms can significantly shorten the pulses
necessary to produce a given unitary, even for a single rotation, as demonstrated in \cite{castro_optimal_2022}.
If one then tests those optimized pulses in the presence
of decoherence (QOCT-S-L, orange curves), the errors grow as
expected. We observe a minimum in the curve (indicated with a marker), a maximum-fidelity sweet spot where neither control
nor decoherence errors are dominant with respect to each other. This is precisely the compromise mentioned above,
attained when simultaneously minimizing the two different errors that
are present in the process.

The results obtained from QOCT-L-L (green curves) are the best in the presence of
decoherence. The difference with
respect to the QOCT-S-L fidelities is larger at longer durations $T$,
and becomes negligible at small $T$. This is not surprising, since QOCT-S and QOCT-L
are only effectively different if decoherence is relevant, i.e. for large durations. 
For the same reason, the difference between QOCT-S and QOCT-L is significantly larger for smaller $T_2$ (top panel), but smaller at larger $T_2$ (bottom panel), i.e. if the dephasing is weaker. These QOCT-L curves also have a sweet spot, for the same reasons as before. This minimum infidelity is lower than the one in the QOCT-S-L curve, which shows how the QOCT-L-L procedure helps to obtain the pulse with the minimum possible error.

As the complexity of the target unitary increases, from $U_1$ to $U_4$
and then to $U_\text{Toffoli}$, we observe two main effects:  First, the
minimum duration required for the monochromatic sequences grows (as
more rotations are needed): the M-S and M-L curves are not visible at these durations for either target unitary, as
$T_{\rm min}$ is too high.

And second, the difference between QOCT-S-L and QOCT-L-L becomes
larger as the complexity of the gate grows. These are all consequences of the
longer durations that are necessary to implement the operations. As a final comment regarding
Fig.~\ref{fig:iF_T}: notice how the QOCT
curves are not very smooth; the reason is that the algorithm may not necessarily find
the global optimum. This difficulty probably grows with increasing gate complexity.

Fig.~\ref{fig:pulses} shows some of the control pulses obtained with QOCT,
together with the absolute value of their Fourier transforms,
computed for the Toffoli gate, using $T_2 = 50\tau$ and $T\approx 7.68\tau$.
Panels (a) and (b) display the pulse obtained with QOCT-S, whereas panels (c) and (d) display the
pulse obtained with QOCT-L.  One can see how the designed parametrization of the pulses fixes
the maximum amplitude for both pulses, that cannot surpass 10~mT. 
Likewise, the frequencies above $\omega_{\rm max}$ are negligible,
as it can be seen in the right-side panels (b) and (d), due
to the chosen parametrization and to the use of a penalty on higher frequencies.
	
\begin{figure}
    \centering
    \includegraphics[width = \linewidth]{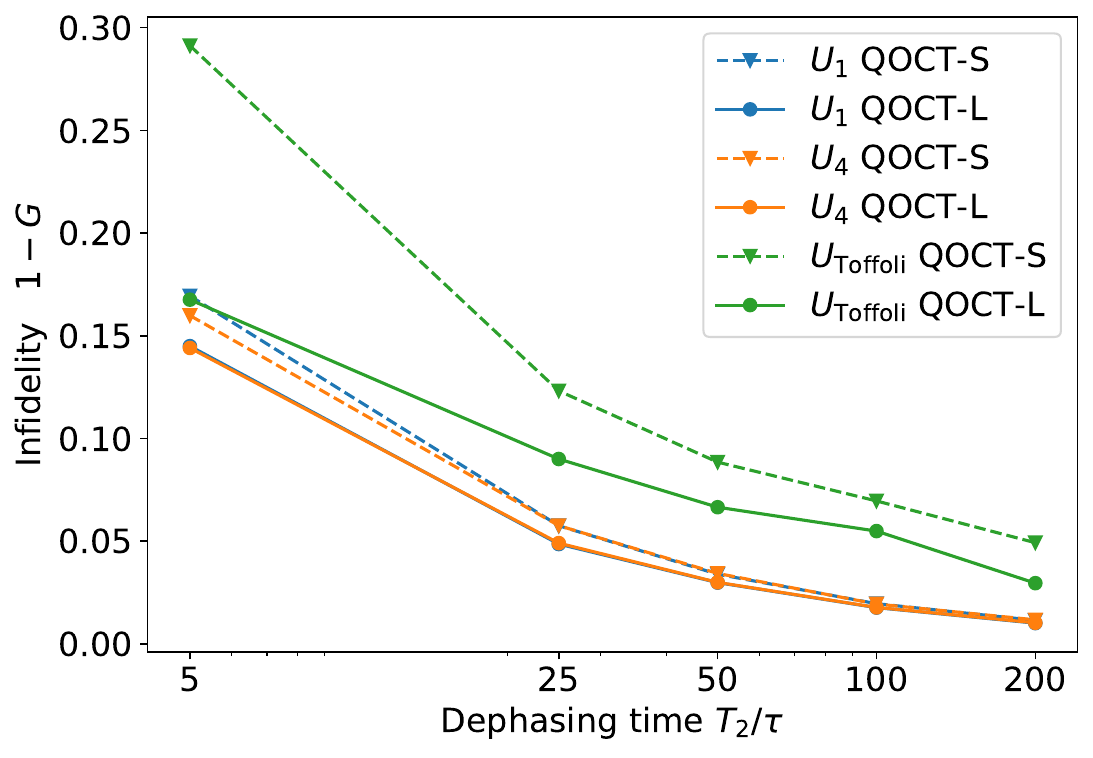}
    \caption{Minimum infidelities achieved by the QOCT-S (triangle marker, dashed lines) and QOCT-L (circular marker, solid lines) methods for all three target unitaries, as a function of the dephasing time $T_2$. The lines are guides to the eye.}
    \label{fig:iF_T2}
\end{figure}
	
Finally, Fig.~\ref{fig:iF_T2} shows the minimum infidelities achieved for
each gate using the QOCT-S and QOCT-L methods, as a function of
$T_2$. This plot summarizes the previous results, as it shows the smallest possible
error that can be obtained with each method.
It can be seen how, when the dephasing is weak (long $T_2$), both methods perform comparably
well. However, in the presence of strong dephasing, the QOCT-L method achieves
substantially lower errors. This difference is also larger for the most complex of the unitaries, $U_\text{Toffoli}$.

\section{\label{sec:conclusions_outlook}Conclusions and Outlook}

This work investigates the use of QOCT for open quantum systems to
find pulses capable of implementing operations in molecular
nanomagnets subject to strong dephasing. Our results show that
accounting for decoherence when numerically designing control pulses
makes a significant difference. Monochromatic pulse sequences show a
bad performance in the presence of pure dephasing, and so do the
waveforms optimized using QOCT on top of Schrödinger's
equation. However, modelling dephasing using Lindblad's master
equation and applying QOCT to it yields higher gate fidelities. The
difference is larger the shorter the coherence times and the more
complex the target operations.

This tool is useful for two reasons. First, it allows one to give a
realistic numerical bound to the achievable gate fidelities on these
systems, in the presence of dephasing and accounting for the amplitude
and sampling-rate limitations of the instrumentation. And second, it
finds the optimal waveforms for this purpose, outperforming the other
approaches.

We consider extending this method in several directions. First, one could add amplitude damping ($T_1$) to the model, to better account for all possible sources of errors. Although we do not expect any significant differences in the case of our platform, in which dephasing is orders of magnitude stronger, this could be useful in other circumstances. Second, we also consider the use of more realistic master equations -- for example, employing time-dependent Lindblad operators. We have ignored this possibiligy so far because, up to first order of approximation, the dissipative operators are independent of the drive, and therefore invariant in time. However, the use of large amplitudes and short pulses may imply a regime where second and higher order terms are not negligible, and there is a relevant time-dependent contribution to the dissipator.

\begin{acknowledgments}
    The authors would like to thank D. Zueco for his support with the theory of open quantum systems.

    This work has received support from grants TED2021-131447B-C21, PID2022-140923NB-C21, and PID2021-123251NB-I00 funded by MCIN/AEI/10.13039/501100011033, ERDF 'A way of making Europe' and ESF 'Investing in your future', and from the Gobierno de Arag\'on grant E09-17R-Q-MAD. This study forms also part of the Advanced Materials and Quantum Communication programmes, supported by MCIN with funding from European Union NextGenerationEU (PRTR-C17.I1), by Gobierno de Arag\'on, and by CSIC (PTI001).
\end{acknowledgments}

\appendix

\section{\label{appendix:phi} {Parametrization and penalty on the control function}}

As mentioned above, the parameterized form of the control functions that we use is
$f(u, t) = \Phi(\tilde{f}(u, t))$,
where $\tilde{f}$ is the Fourier expansion~(\ref{eq:Fourierexpansion}).
In order to bound the amplitude of the signal, this expansion
is modulated by a function $\Phi$ defined in such a way that:
\begin{equation}
	\vert f(u, t)\vert \le \kappa\quad\textrm{ at any $t$.}
\end{equation}
In addition, we require that:
(1) for $\vert 
x\vert \ll \kappa$, $\Phi(x) \approx x$ so that, for low 
amplitudes, the Fourier series is not altered;
(2) as $\vert x\vert$ increases and approaches $\kappa$, the function should 
smoothly evolve towards the constant $\Phi(x) = \kappa$ (or $\Phi(x) = -\kappa$ 
if  $x<0$)
There exist different choices for $\Phi$ that can fulfill these
conditions. The function that we have used here is:
\begin{equation}
\Phi(x) = \left\{
\begin{array}{l}
x\quad \textrm{ if }0\le x \le \frac{3}{4}\kappa
\\
\kappa\quad \textrm{ if }x \ge \kappa + \frac{1}{4}\kappa
\\
\textrm{an Akima cubic spline interpolation~\cite{Akima1970}}
\\ \quad \textrm{if } x \in (\frac{3}{4}\kappa, \kappa + \frac{1}{4}\kappa)
\end{array}
\right.
\end{equation}
This function is antisymmetric: $\Phi(-x) = - \Phi(x)$.

This modulation will slightly distort the original frequency spectrum of $\tilde{f}$, and in particular it
may lead to a violation of the frequency bound $\omega_\text{max}$.
Therefore, this parameterization guarantees the amplitude limit, but only
approximately limits the bandwidth limit of the control field.
For this reason, we also use a penalty function in the definition
of the target function (see Eq.~(\ref{eq:multitarget_fidelity_new})), defined as:
\begin{equation}
P(u) = - \alpha \sum_{\omega_k > \omega_{\rm max}} \vert \mathcal{F}_{k}(u)\vert^2\,,
\end{equation}
where $\mathcal{F}_k$ is the $k$-th component of the Fourier series of $f$,
$\omega_k = \frac{2\pi}{T}k$, and $\omega_{\rm max}$ is the cutoff frequency
orginally set by the expansion \ref{eq:Fourierexpansion}. The constant $\alpha > 0$ decides
the weight of the penalty function. By adding this function to the target functional,
the search algorithm will avoid the frequencies above $\omega_{\rm max}$.


\bibliography{main}

\begin{thebibliography}{53}%
\makeatletter
\providecommand \@ifxundefined [1]{%
 \@ifx{#1\undefined}
}%
\providecommand \@ifnum [1]{%
 \ifnum #1\expandafter \@firstoftwo
 \else \expandafter \@secondoftwo
 \fi
}%
\providecommand \@ifx [1]{%
 \ifx #1\expandafter \@firstoftwo
 \else \expandafter \@secondoftwo
 \fi
}%
\providecommand \natexlab [1]{#1}%
\providecommand \enquote  [1]{``#1''}%
\providecommand \bibnamefont  [1]{#1}%
\providecommand \bibfnamefont [1]{#1}%
\providecommand \citenamefont [1]{#1}%
\providecommand \href@noop [0]{\@secondoftwo}%
\providecommand \href [0]{\begingroup \@sanitize@url \@href}%
\providecommand \@href[1]{\@@startlink{#1}\@@href}%
\providecommand \@@href[1]{\endgroup#1\@@endlink}%
\providecommand \@sanitize@url [0]{\catcode `\\12\catcode `\$12\catcode
  `\&12\catcode `\#12\catcode `\^12\catcode `\_12\catcode `\%12\relax}%
\providecommand \@@startlink[1]{}%
\providecommand \@@endlink[0]{}%
\providecommand \url  [0]{\begingroup\@sanitize@url \@url }%
\providecommand \@url [1]{\endgroup\@href {#1}{\urlprefix }}%
\providecommand \urlprefix  [0]{URL }%
\providecommand \Eprint [0]{\href }%
\providecommand \doibase [0]{https://doi.org/}%
\providecommand \selectlanguage [0]{\@gobble}%
\providecommand \bibinfo  [0]{\@secondoftwo}%
\providecommand \bibfield  [0]{\@secondoftwo}%
\providecommand \translation [1]{[#1]}%
\providecommand \BibitemOpen [0]{}%
\providecommand \bibitemStop [0]{}%
\providecommand \bibitemNoStop [0]{.\EOS\space}%
\providecommand \EOS [0]{\spacefactor3000\relax}%
\providecommand \BibitemShut  [1]{\csname bibitem#1\endcsname}%
\let\auto@bib@innerbib\@empty
\bibitem [{\citenamefont {Knill}\ and\ \citenamefont
  {Laflamme}(1997)}]{knill_theory_1997}%
  \BibitemOpen
  \bibfield  {author} {\bibinfo {author} {\bibfnamefont {E.}~\bibnamefont
  {Knill}}\ and\ \bibinfo {author} {\bibfnamefont {R.}~\bibnamefont
  {Laflamme}},\ }\bibfield  {title} {\bibinfo {title} {Theory of quantum
  error-correcting codes},\ }\href
  {https://link.aps.org/doi/10.1103/PhysRevA.55.900} {\bibfield  {journal}
  {\bibinfo  {journal} {Phys. Rev. A}\ }\textbf {\bibinfo {volume} {55}},\
  \bibinfo {pages} {900} (\bibinfo {year} {1997})}\BibitemShut {NoStop}%
\bibitem [{\citenamefont {Postler}\ \emph {et~al.}(2022)\citenamefont
  {Postler}, \citenamefont {Heu{\ss}en}, \citenamefont {Pogorelov},
  \citenamefont {Rispler}, \citenamefont {Feldker}, \citenamefont {Meth},
  \citenamefont {Marciniak}, \citenamefont {Stricker}, \citenamefont
  {Ringbauer}, \citenamefont {Blatt}, \citenamefont {Schindler}, \citenamefont
  {Müller},\ and\ \citenamefont {Monz}}]{postler_demonstration_2022}%
  \BibitemOpen
  \bibfield  {author} {\bibinfo {author} {\bibfnamefont {L.}~\bibnamefont
  {Postler}}, \bibinfo {author} {\bibfnamefont {S.}~\bibnamefont {Heu{\ss}en}},
  \bibinfo {author} {\bibfnamefont {I.}~\bibnamefont {Pogorelov}}, \bibinfo
  {author} {\bibfnamefont {M.}~\bibnamefont {Rispler}}, \bibinfo {author}
  {\bibfnamefont {T.}~\bibnamefont {Feldker}}, \bibinfo {author} {\bibfnamefont
  {M.}~\bibnamefont {Meth}}, \bibinfo {author} {\bibfnamefont {C.~D.}\
  \bibnamefont {Marciniak}}, \bibinfo {author} {\bibfnamefont {R.}~\bibnamefont
  {Stricker}}, \bibinfo {author} {\bibfnamefont {M.}~\bibnamefont {Ringbauer}},
  \bibinfo {author} {\bibfnamefont {R.}~\bibnamefont {Blatt}}, \bibinfo
  {author} {\bibfnamefont {P.}~\bibnamefont {Schindler}}, \bibinfo {author}
  {\bibfnamefont {M.}~\bibnamefont {Müller}},\ and\ \bibinfo {author}
  {\bibfnamefont {T.}~\bibnamefont {Monz}},\ }\bibfield  {title} {\bibinfo
  {title} {Demonstration of fault-tolerant universal quantum gate operations},\
  }\href {https://www.nature.com/articles/s41586-022-04721-1} {\bibfield
  {journal} {\bibinfo  {journal} {Nature}\ }\textbf {\bibinfo {volume} {605}},\
  \bibinfo {pages} {675} (\bibinfo {year} {2022})}\BibitemShut {NoStop}%
\bibitem [{\citenamefont {Bluvstein}\ \emph {et~al.}(2022)\citenamefont
  {Bluvstein}, \citenamefont {Levine}, \citenamefont {Semeghini}, \citenamefont
  {Wang}, \citenamefont {Ebadi}, \citenamefont {Kalinowski}, \citenamefont
  {Keesling}, \citenamefont {Maskara}, \citenamefont {Pichler}, \citenamefont
  {Greiner}, \citenamefont {Vuletić},\ and\ \citenamefont
  {Lukin}}]{bluvstein_quantum_2022}%
  \BibitemOpen
  \bibfield  {author} {\bibinfo {author} {\bibfnamefont {D.}~\bibnamefont
  {Bluvstein}}, \bibinfo {author} {\bibfnamefont {H.}~\bibnamefont {Levine}},
  \bibinfo {author} {\bibfnamefont {G.}~\bibnamefont {Semeghini}}, \bibinfo
  {author} {\bibfnamefont {T.~T.}\ \bibnamefont {Wang}}, \bibinfo {author}
  {\bibfnamefont {S.}~\bibnamefont {Ebadi}}, \bibinfo {author} {\bibfnamefont
  {M.}~\bibnamefont {Kalinowski}}, \bibinfo {author} {\bibfnamefont
  {A.}~\bibnamefont {Keesling}}, \bibinfo {author} {\bibfnamefont
  {N.}~\bibnamefont {Maskara}}, \bibinfo {author} {\bibfnamefont
  {H.}~\bibnamefont {Pichler}}, \bibinfo {author} {\bibfnamefont
  {M.}~\bibnamefont {Greiner}}, \bibinfo {author} {\bibfnamefont
  {V.}~\bibnamefont {Vuletić}},\ and\ \bibinfo {author} {\bibfnamefont
  {M.~D.}\ \bibnamefont {Lukin}},\ }\bibfield  {title} {\bibinfo {title} {A
  quantum processor based on coherent transport of entangled atom arrays},\
  }\href {https://www.nature.com/articles/s41586-022-04592-6} {\bibfield
  {journal} {\bibinfo  {journal} {Nature}\ }\textbf {\bibinfo {volume} {604}},\
  \bibinfo {pages} {451} (\bibinfo {year} {2022})}\BibitemShut {NoStop}%
\bibitem [{\citenamefont {Abobeih}\ \emph {et~al.}(2022)\citenamefont
  {Abobeih}, \citenamefont {Wang}, \citenamefont {Randall}, \citenamefont
  {Loenen}, \citenamefont {Bradley}, \citenamefont {Markham}, \citenamefont
  {Twitchen}, \citenamefont {Terhal},\ and\ \citenamefont
  {Taminiau}}]{abobeih_fault-tolerant_2022}%
  \BibitemOpen
  \bibfield  {author} {\bibinfo {author} {\bibfnamefont {M.~H.}\ \bibnamefont
  {Abobeih}}, \bibinfo {author} {\bibfnamefont {Y.}~\bibnamefont {Wang}},
  \bibinfo {author} {\bibfnamefont {J.}~\bibnamefont {Randall}}, \bibinfo
  {author} {\bibfnamefont {S.~J.~H.}\ \bibnamefont {Loenen}}, \bibinfo {author}
  {\bibfnamefont {C.~E.}\ \bibnamefont {Bradley}}, \bibinfo {author}
  {\bibfnamefont {M.}~\bibnamefont {Markham}}, \bibinfo {author} {\bibfnamefont
  {D.~J.}\ \bibnamefont {Twitchen}}, \bibinfo {author} {\bibfnamefont {B.~M.}\
  \bibnamefont {Terhal}},\ and\ \bibinfo {author} {\bibfnamefont {T.~H.}\
  \bibnamefont {Taminiau}},\ }\bibfield  {title} {\bibinfo {title}
  {Fault-tolerant operation of a logical qubit in a diamond quantum
  processor},\ }\href {https://www.nature.com/articles/s41586-022-04819-6}
  {\bibfield  {journal} {\bibinfo  {journal} {Nature}\ }\textbf {\bibinfo
  {volume} {606}},\ \bibinfo {pages} {884} (\bibinfo {year}
  {2022})}\BibitemShut {NoStop}%
\bibitem [{\citenamefont {Takeda}\ \emph {et~al.}(2022)\citenamefont {Takeda},
  \citenamefont {Noiri}, \citenamefont {Nakajima}, \citenamefont {Kobayashi},\
  and\ \citenamefont {Tarucha}}]{takeda_quantum_2022}%
  \BibitemOpen
  \bibfield  {author} {\bibinfo {author} {\bibfnamefont {K.}~\bibnamefont
  {Takeda}}, \bibinfo {author} {\bibfnamefont {A.}~\bibnamefont {Noiri}},
  \bibinfo {author} {\bibfnamefont {T.}~\bibnamefont {Nakajima}}, \bibinfo
  {author} {\bibfnamefont {T.}~\bibnamefont {Kobayashi}},\ and\ \bibinfo
  {author} {\bibfnamefont {S.}~\bibnamefont {Tarucha}},\ }\bibfield  {title}
  {\bibinfo {title} {Quantum error correction with silicon spin qubits},\
  }\href {https://www.nature.com/articles/s41586-022-04986-6} {\bibfield
  {journal} {\bibinfo  {journal} {Nature}\ }\textbf {\bibinfo {volume} {608}},\
  \bibinfo {pages} {682} (\bibinfo {year} {2022})}\BibitemShut {NoStop}%
\bibitem [{\citenamefont {{\relax Google Quantum
  AI}}(2023)}]{acharya_suppressing_2023}%
  \BibitemOpen
  \bibfield  {author} {\bibinfo {author} {\bibnamefont {{\relax Google Quantum
  AI}}},\ }\bibfield  {title} {\bibinfo {title} {Suppressing quantum errors by
  scaling a surface code logical qubit},\ }\href
  {https://www.nature.com/articles/s41586-022-05434-1} {\bibfield  {journal}
  {\bibinfo  {journal} {Nature}\ }\textbf {\bibinfo {volume} {614}},\ \bibinfo
  {pages} {676} (\bibinfo {year} {2023})}\BibitemShut {NoStop}%
\bibitem [{\citenamefont {Gottesman}\ \emph {et~al.}(2001)\citenamefont
  {Gottesman}, \citenamefont {Kitaev},\ and\ \citenamefont
  {Preskill}}]{gottesman_encoding_2001}%
  \BibitemOpen
  \bibfield  {author} {\bibinfo {author} {\bibfnamefont {D.}~\bibnamefont
  {Gottesman}}, \bibinfo {author} {\bibfnamefont {A.}~\bibnamefont {Kitaev}},\
  and\ \bibinfo {author} {\bibfnamefont {J.}~\bibnamefont {Preskill}},\
  }\bibfield  {title} {\bibinfo {title} {Encoding a qubit in an oscillator},\
  }\href {https://link.aps.org/doi/10.1103/PhysRevA.64.012310} {\bibfield
  {journal} {\bibinfo  {journal} {Phys. Rev. A}\ }\textbf {\bibinfo {volume}
  {64}},\ \bibinfo {pages} {012310} (\bibinfo {year} {2001})}\BibitemShut
  {NoStop}%
\bibitem [{\citenamefont {Leuenberger}\ and\ \citenamefont
  {Loss}(2001)}]{leuenberger_quantum_2001}%
  \BibitemOpen
  \bibfield  {author} {\bibinfo {author} {\bibfnamefont {M.~N.}\ \bibnamefont
  {Leuenberger}}\ and\ \bibinfo {author} {\bibfnamefont {D.}~\bibnamefont
  {Loss}},\ }\bibfield  {title} {\bibinfo {title} {Quantum computing in
  molecular magnets},\ }\href {https://doi.org/10.1038/35071024} {\bibfield
  {journal} {\bibinfo  {journal} {Nature}\ }\textbf {\bibinfo {volume} {410}},\
  \bibinfo {pages} {789} (\bibinfo {year} {2001})}\BibitemShut {NoStop}%
\bibitem [{\citenamefont {Brennen}\ \emph {et~al.}(2005)\citenamefont
  {Brennen}, \citenamefont {O'Leary},\ and\ \citenamefont
  {Bullock}}]{brennen_criteria_2005}%
  \BibitemOpen
  \bibfield  {author} {\bibinfo {author} {\bibfnamefont {G.~K.}\ \bibnamefont
  {Brennen}}, \bibinfo {author} {\bibfnamefont {D.~P.}\ \bibnamefont
  {O'Leary}},\ and\ \bibinfo {author} {\bibfnamefont {S.~S.}\ \bibnamefont
  {Bullock}},\ }\bibfield  {title} {\bibinfo {title} {Criteria for exact qudit
  universality},\ }\href {https://link.aps.org/doi/10.1103/PhysRevA.71.052318}
  {\bibfield  {journal} {\bibinfo  {journal} {Phys. Rev. A}\ }\textbf {\bibinfo
  {volume} {71}},\ \bibinfo {pages} {052318} (\bibinfo {year}
  {2005})}\BibitemShut {NoStop}%
\bibitem [{\citenamefont {Lanyon}\ \emph {et~al.}(2009)\citenamefont {Lanyon},
  \citenamefont {Barbieri}, \citenamefont {Almeida}, \citenamefont {Jennewein},
  \citenamefont {Ralph}, \citenamefont {Resch}, \citenamefont {Pryde},
  \citenamefont {O’Brien}, \citenamefont {Alexei},\ and\ \citenamefont
  {White}}]{lanyon_simplifying_2009}%
  \BibitemOpen
  \bibfield  {author} {\bibinfo {author} {\bibfnamefont {B.~P.}\ \bibnamefont
  {Lanyon}}, \bibinfo {author} {\bibfnamefont {M.}~\bibnamefont {Barbieri}},
  \bibinfo {author} {\bibfnamefont {M.~P.}\ \bibnamefont {Almeida}}, \bibinfo
  {author} {\bibfnamefont {T.}~\bibnamefont {Jennewein}}, \bibinfo {author}
  {\bibfnamefont {T.~C.}\ \bibnamefont {Ralph}}, \bibinfo {author}
  {\bibfnamefont {K.~J.}\ \bibnamefont {Resch}}, \bibinfo {author}
  {\bibfnamefont {G.~J.}\ \bibnamefont {Pryde}}, \bibinfo {author}
  {\bibfnamefont {J.~L.}\ \bibnamefont {O’Brien}}, \bibinfo {author}
  {\bibfnamefont {G.}~\bibnamefont {Alexei}},\ and\ \bibinfo {author}
  {\bibfnamefont {A.~G.}\ \bibnamefont {White}},\ }\bibfield  {title} {\bibinfo
  {title} {Simplifying quantum logic using higher-dimensional hilbert spaces},\
  }\href {https://doi.org/10.1038/nphys1150} {\bibfield  {journal} {\bibinfo
  {journal} {Nat. Phys.}\ }\textbf {\bibinfo {volume} {5}},\ \bibinfo {pages}
  {134} (\bibinfo {year} {2009})}\BibitemShut {NoStop}%
\bibitem [{\citenamefont {Kiktenko}\ \emph {et~al.}(2015)\citenamefont
  {Kiktenko}, \citenamefont {Fedorov}, \citenamefont {Strakhov},\ and\
  \citenamefont {Man'ko}}]{kiktenko_single_2015}%
  \BibitemOpen
  \bibfield  {author} {\bibinfo {author} {\bibfnamefont {E.~O.}\ \bibnamefont
  {Kiktenko}}, \bibinfo {author} {\bibfnamefont {A.~K.}\ \bibnamefont
  {Fedorov}}, \bibinfo {author} {\bibfnamefont {A.~A.}\ \bibnamefont
  {Strakhov}},\ and\ \bibinfo {author} {\bibfnamefont {V.~I.}\ \bibnamefont
  {Man'ko}},\ }\bibfield  {title} {\bibinfo {title} {Single qudit realization
  of the deutsch algorithm using superconducting many-level quantum circuits},\
  }\href {http://www.sciencedirect.com/science/article/pii/S0375960115002753}
  {\bibfield  {journal} {\bibinfo  {journal} {Phys. Lett. A}\ }\textbf
  {\bibinfo {volume} {379}},\ \bibinfo {pages} {1409 } (\bibinfo {year}
  {2015})}\BibitemShut {NoStop}%
\bibitem [{\citenamefont {Pirandola}\ \emph {et~al.}(2008)\citenamefont
  {Pirandola}, \citenamefont {Mancini}, \citenamefont {Braunstein},\ and\
  \citenamefont {Vitali}}]{pirandola_minimal_2008}%
  \BibitemOpen
  \bibfield  {author} {\bibinfo {author} {\bibfnamefont {S.}~\bibnamefont
  {Pirandola}}, \bibinfo {author} {\bibfnamefont {S.}~\bibnamefont {Mancini}},
  \bibinfo {author} {\bibfnamefont {S.~L.}\ \bibnamefont {Braunstein}},\ and\
  \bibinfo {author} {\bibfnamefont {D.}~\bibnamefont {Vitali}},\ }\bibfield
  {title} {\bibinfo {title} {Minimal qudit code for a qubit in the
  phase-damping channel},\ }\href
  {https://link.aps.org/doi/10.1103/PhysRevA.77.032309} {\bibfield  {journal}
  {\bibinfo  {journal} {Phys. Rev. A}\ }\textbf {\bibinfo {volume} {77}},\
  \bibinfo {pages} {032309} (\bibinfo {year} {2008})}\BibitemShut {NoStop}%
\bibitem [{\citenamefont {Chiesa}\ \emph {et~al.}(2020)\citenamefont {Chiesa},
  \citenamefont {Macaluso}, \citenamefont {Petiziol}, \citenamefont
  {Wimberger}, \citenamefont {Santini},\ and\ \citenamefont
  {Carretta}}]{chiesa_molecular_2020}%
  \BibitemOpen
  \bibfield  {author} {\bibinfo {author} {\bibfnamefont {A.}~\bibnamefont
  {Chiesa}}, \bibinfo {author} {\bibfnamefont {E.}~\bibnamefont {Macaluso}},
  \bibinfo {author} {\bibfnamefont {F.}~\bibnamefont {Petiziol}}, \bibinfo
  {author} {\bibfnamefont {S.}~\bibnamefont {Wimberger}}, \bibinfo {author}
  {\bibfnamefont {P.}~\bibnamefont {Santini}},\ and\ \bibinfo {author}
  {\bibfnamefont {S.}~\bibnamefont {Carretta}},\ }\bibfield  {title} {\bibinfo
  {title} {Molecular nanomagnets as qubits with embedded quantum-error
  correction},\ }\href {https://doi.org/10.1021/acs.jpclett.0c02213} {\bibfield
   {journal} {\bibinfo  {journal} {J. Phys. Chem. Lett.}\ }\textbf {\bibinfo
  {volume} {11}},\ \bibinfo {pages} {8610} (\bibinfo {year}
  {2020})}\BibitemShut {NoStop}%
\bibitem [{\citenamefont {Lapkiewicz}\ \emph {et~al.}(2011)\citenamefont
  {Lapkiewicz}, \citenamefont {Li}, \citenamefont {Schaeff}, \citenamefont
  {Langford}, \citenamefont {Ramelow}, \citenamefont {Wieśniak},\ and\
  \citenamefont {Zeilinger}}]{lapkiewicz_experimental_2011}%
  \BibitemOpen
  \bibfield  {author} {\bibinfo {author} {\bibfnamefont {R.}~\bibnamefont
  {Lapkiewicz}}, \bibinfo {author} {\bibfnamefont {P.}~\bibnamefont {Li}},
  \bibinfo {author} {\bibfnamefont {C.}~\bibnamefont {Schaeff}}, \bibinfo
  {author} {\bibfnamefont {N.~K.}\ \bibnamefont {Langford}}, \bibinfo {author}
  {\bibfnamefont {S.}~\bibnamefont {Ramelow}}, \bibinfo {author} {\bibfnamefont
  {M.}~\bibnamefont {Wieśniak}},\ and\ \bibinfo {author} {\bibfnamefont
  {A.}~\bibnamefont {Zeilinger}},\ }\bibfield  {title} {\bibinfo {title}
  {Experimental non-classicality of an indivisible quantum system},\ }\href
  {https://doi.org/10.1038/nature10119} {\bibfield  {journal} {\bibinfo
  {journal} {Nature}\ }\textbf {\bibinfo {volume} {474}},\ \bibinfo {pages}
  {490} (\bibinfo {year} {2011})}\BibitemShut {NoStop}%
\bibitem [{\citenamefont {Ringbauer}\ \emph {et~al.}(2022)\citenamefont
  {Ringbauer}, \citenamefont {Meth}, \citenamefont {Postler}, \citenamefont
  {Stricker}, \citenamefont {Blatt}, \citenamefont {Schindler},\ and\
  \citenamefont {Monz}}]{ringbauer_universal_2022}%
  \BibitemOpen
  \bibfield  {author} {\bibinfo {author} {\bibfnamefont {M.}~\bibnamefont
  {Ringbauer}}, \bibinfo {author} {\bibfnamefont {M.}~\bibnamefont {Meth}},
  \bibinfo {author} {\bibfnamefont {L.}~\bibnamefont {Postler}}, \bibinfo
  {author} {\bibfnamefont {R.}~\bibnamefont {Stricker}}, \bibinfo {author}
  {\bibfnamefont {R.}~\bibnamefont {Blatt}}, \bibinfo {author} {\bibfnamefont
  {P.}~\bibnamefont {Schindler}},\ and\ \bibinfo {author} {\bibfnamefont
  {T.}~\bibnamefont {Monz}},\ }\bibfield  {title} {\bibinfo {title} {A
  universal qudit quantum processor with trapped ions},\ }\href
  {https://www.nature.com/articles/s41567-022-01658-0} {\bibfield  {journal}
  {\bibinfo  {journal} {Nat. Phys.}\ }\textbf {\bibinfo {volume} {18}},\
  \bibinfo {pages} {1053} (\bibinfo {year} {2022})}\BibitemShut {NoStop}%
\bibitem [{\citenamefont {Asaad}\ \emph {et~al.}(2020)\citenamefont {Asaad},
  \citenamefont {Mourik}, \citenamefont {Joecker}, \citenamefont {Johnson},
  \citenamefont {Baczewski}, \citenamefont {Firgau}, \citenamefont {Madzik},
  \citenamefont {Schmitt}, \citenamefont {Pla}, \citenamefont {Hudson},
  \citenamefont {Itoh}, \citenamefont {McCallum}, \citenamefont {Dzurak},
  \citenamefont {Laucht},\ and\ \citenamefont {Morello}}]{asaad_coherent_2020}%
  \BibitemOpen
  \bibfield  {author} {\bibinfo {author} {\bibfnamefont {S.}~\bibnamefont
  {Asaad}}, \bibinfo {author} {\bibfnamefont {V.}~\bibnamefont {Mourik}},
  \bibinfo {author} {\bibfnamefont {B.}~\bibnamefont {Joecker}}, \bibinfo
  {author} {\bibfnamefont {M.~A.~I.}\ \bibnamefont {Johnson}}, \bibinfo
  {author} {\bibfnamefont {A.~D.}\ \bibnamefont {Baczewski}}, \bibinfo {author}
  {\bibfnamefont {H.~R.}\ \bibnamefont {Firgau}}, \bibinfo {author}
  {\bibfnamefont {M.~T.}\ \bibnamefont {Madzik}}, \bibinfo {author}
  {\bibfnamefont {V.}~\bibnamefont {Schmitt}}, \bibinfo {author} {\bibfnamefont
  {J.~J.}\ \bibnamefont {Pla}}, \bibinfo {author} {\bibfnamefont {F.~E.}\
  \bibnamefont {Hudson}}, \bibinfo {author} {\bibfnamefont {K.~M.}\
  \bibnamefont {Itoh}}, \bibinfo {author} {\bibfnamefont {J.~C.}\ \bibnamefont
  {McCallum}}, \bibinfo {author} {\bibfnamefont {A.~S.}\ \bibnamefont
  {Dzurak}}, \bibinfo {author} {\bibfnamefont {A.}~\bibnamefont {Laucht}},\
  and\ \bibinfo {author} {\bibfnamefont {A.}~\bibnamefont {Morello}},\
  }\bibfield  {title} {\bibinfo {title} {Coherent electrical control of a
  single high-spin nucleus in silicon},\ }\href
  {https://doi.org/10.1038/s41586-020-2057-7} {\bibfield  {journal} {\bibinfo
  {journal} {Nature}\ }\textbf {\bibinfo {volume} {579}},\ \bibinfo {pages}
  {205} (\bibinfo {year} {2020})}\BibitemShut {NoStop}%
\bibitem [{\citenamefont {Neeley}\ \emph {et~al.}(2009)\citenamefont {Neeley},
  \citenamefont {Ansmann}, \citenamefont {Bialczak}, \citenamefont {Hofheinz},
  \citenamefont {Lucero}, \citenamefont {O'Connell}, \citenamefont {Sank},
  \citenamefont {Wang}, \citenamefont {Wenner}, \citenamefont {Cleland},
  \citenamefont {Geller},\ and\ \citenamefont
  {Martinis}}]{neeley_emulation_2009}%
  \BibitemOpen
  \bibfield  {author} {\bibinfo {author} {\bibfnamefont {M.}~\bibnamefont
  {Neeley}}, \bibinfo {author} {\bibfnamefont {M.}~\bibnamefont {Ansmann}},
  \bibinfo {author} {\bibfnamefont {R.~C.}\ \bibnamefont {Bialczak}}, \bibinfo
  {author} {\bibfnamefont {M.}~\bibnamefont {Hofheinz}}, \bibinfo {author}
  {\bibfnamefont {E.}~\bibnamefont {Lucero}}, \bibinfo {author} {\bibfnamefont
  {A.~D.}\ \bibnamefont {O'Connell}}, \bibinfo {author} {\bibfnamefont
  {D.}~\bibnamefont {Sank}}, \bibinfo {author} {\bibfnamefont {H.}~\bibnamefont
  {Wang}}, \bibinfo {author} {\bibfnamefont {J.}~\bibnamefont {Wenner}},
  \bibinfo {author} {\bibfnamefont {A.~N.}\ \bibnamefont {Cleland}}, \bibinfo
  {author} {\bibfnamefont {M.~R.}\ \bibnamefont {Geller}},\ and\ \bibinfo
  {author} {\bibfnamefont {J.~M.}\ \bibnamefont {Martinis}},\ }\bibfield
  {title} {\bibinfo {title} {Emulation of a quantum spin with a superconducting
  phase qudit},\ }\href {https://doi.org/10.1126/science.1173440} {\bibfield
  {journal} {\bibinfo  {journal} {Science}\ }\textbf {\bibinfo {volume}
  {325}},\ \bibinfo {pages} {722} (\bibinfo {year} {2009})}\BibitemShut
  {NoStop}%
\bibitem [{\citenamefont {Brif}\ \emph {et~al.}(2010)\citenamefont {Brif},
  \citenamefont {Chakrabarti},\ and\ \citenamefont
  {Rabitz}}]{brif_control_2010}%
  \BibitemOpen
  \bibfield  {author} {\bibinfo {author} {\bibfnamefont {C.}~\bibnamefont
  {Brif}}, \bibinfo {author} {\bibfnamefont {R.}~\bibnamefont {Chakrabarti}},\
  and\ \bibinfo {author} {\bibfnamefont {H.}~\bibnamefont {Rabitz}},\
  }\bibfield  {title} {\bibinfo {title} {Control of quantum phenomena: past
  present and future},\ }\href@noop {} {\bibfield  {journal} {\bibinfo
  {journal} {New J. Phys.}\ }\textbf {\bibinfo {volume} {12}},\ \bibinfo
  {pages} {075008} (\bibinfo {year} {2010})}\BibitemShut {NoStop}%
\bibitem [{\citenamefont {Glaser}\ \emph {et~al.}(2015)\citenamefont {Glaser},
  \citenamefont {Boscain}, \citenamefont {Calarco}, \citenamefont {Koch},
  \citenamefont {Köckenberger}, \citenamefont {Kosloff}, \citenamefont
  {Kuprov}, \citenamefont {Luy}, \citenamefont {Schirmer}, \citenamefont
  {Schulte-Herbrüggen}, \citenamefont {Sugny},\ and\ \citenamefont
  {Wilhelm}}]{glaser_training_2015}%
  \BibitemOpen
  \bibfield  {author} {\bibinfo {author} {\bibfnamefont {S.~J.}\ \bibnamefont
  {Glaser}}, \bibinfo {author} {\bibfnamefont {U.}~\bibnamefont {Boscain}},
  \bibinfo {author} {\bibfnamefont {T.}~\bibnamefont {Calarco}}, \bibinfo
  {author} {\bibfnamefont {C.~P.}\ \bibnamefont {Koch}}, \bibinfo {author}
  {\bibfnamefont {W.}~\bibnamefont {Köckenberger}}, \bibinfo {author}
  {\bibfnamefont {R.}~\bibnamefont {Kosloff}}, \bibinfo {author} {\bibfnamefont
  {I.}~\bibnamefont {Kuprov}}, \bibinfo {author} {\bibfnamefont
  {B.}~\bibnamefont {Luy}}, \bibinfo {author} {\bibfnamefont {S.}~\bibnamefont
  {Schirmer}}, \bibinfo {author} {\bibfnamefont {T.}~\bibnamefont
  {Schulte-Herbrüggen}}, \bibinfo {author} {\bibfnamefont {D.}~\bibnamefont
  {Sugny}},\ and\ \bibinfo {author} {\bibfnamefont {F.~K.}\ \bibnamefont
  {Wilhelm}},\ }\bibfield  {title} {\bibinfo {title} {Training schrödinger's
  cat: quantum optimal control},\ }\href
  {https://doi.org/10.1140/epjd/e2015-60464-1} {\bibfield  {journal} {\bibinfo
  {journal} {The European Physical Journal D}\ }\textbf {\bibinfo {volume}
  {69}},\ \bibinfo {pages} {279} (\bibinfo {year} {2015})}\BibitemShut
  {NoStop}%
\bibitem [{\citenamefont {Koch}\ \emph {et~al.}(2022)\citenamefont {Koch},
  \citenamefont {Boscain}, \citenamefont {Calarco}, \citenamefont {Dirr},
  \citenamefont {Filipp}, \citenamefont {Glaser}, \citenamefont {Kosloff},
  \citenamefont {Montangero}, \citenamefont {Schulte-Herbr{\"u}ggen},
  \citenamefont {Sugny},\ and\ \citenamefont {Wilhelm}}]{Koch2022}%
  \BibitemOpen
  \bibfield  {author} {\bibinfo {author} {\bibfnamefont {C.~P.}\ \bibnamefont
  {Koch}}, \bibinfo {author} {\bibfnamefont {U.}~\bibnamefont {Boscain}},
  \bibinfo {author} {\bibfnamefont {T.}~\bibnamefont {Calarco}}, \bibinfo
  {author} {\bibfnamefont {G.}~\bibnamefont {Dirr}}, \bibinfo {author}
  {\bibfnamefont {S.}~\bibnamefont {Filipp}}, \bibinfo {author} {\bibfnamefont
  {S.~J.}\ \bibnamefont {Glaser}}, \bibinfo {author} {\bibfnamefont
  {R.}~\bibnamefont {Kosloff}}, \bibinfo {author} {\bibfnamefont
  {S.}~\bibnamefont {Montangero}}, \bibinfo {author} {\bibfnamefont
  {T.}~\bibnamefont {Schulte-Herbr{\"u}ggen}}, \bibinfo {author} {\bibfnamefont
  {D.}~\bibnamefont {Sugny}},\ and\ \bibinfo {author} {\bibfnamefont {F.~K.}\
  \bibnamefont {Wilhelm}},\ }\bibfield  {title} {\bibinfo {title} {{Quantum
  Optimal Control in Quantum Technologies. Strategic Report on Current Status,
  Visions and Goals for Research in Europe}},\ }\href
  {https://doi.org/10.1140/epjqt/s40507-022-00138-x} {\bibfield  {journal}
  {\bibinfo  {journal} {EPJ Quantum Technology}\ }\textbf {\bibinfo {volume}
  {9}},\ \bibinfo {pages} {19} (\bibinfo {year} {2022})}\BibitemShut {NoStop}%
\bibitem [{\citenamefont {Palao}\ and\ \citenamefont
  {Kosloff}(2002)}]{Palao2002}%
  \BibitemOpen
  \bibfield  {author} {\bibinfo {author} {\bibfnamefont {J.~P.}\ \bibnamefont
  {Palao}}\ and\ \bibinfo {author} {\bibfnamefont {R.}~\bibnamefont
  {Kosloff}},\ }\bibfield  {title} {\bibinfo {title} {Quantum computing by an
  optimal control algorithm for unitary transformations},\ }\href
  {https://doi.org/10.1103/PhysRevLett.89.188301} {\bibfield  {journal}
  {\bibinfo  {journal} {Phys. Rev. Lett.}\ }\textbf {\bibinfo {volume} {89}},\
  \bibinfo {pages} {188301} (\bibinfo {year} {2002})}\BibitemShut {NoStop}%
\bibitem [{\citenamefont {Palao}\ and\ \citenamefont
  {Kosloff}(2003)}]{Palao2003}%
  \BibitemOpen
  \bibfield  {author} {\bibinfo {author} {\bibfnamefont {J.~P.}\ \bibnamefont
  {Palao}}\ and\ \bibinfo {author} {\bibfnamefont {R.}~\bibnamefont
  {Kosloff}},\ }\bibfield  {title} {\bibinfo {title} {Optimal control theory
  for unitary transformations},\ }\href
  {https://doi.org/10.1103/PhysRevA.68.062308} {\bibfield  {journal} {\bibinfo
  {journal} {Phys. Rev. A}\ }\textbf {\bibinfo {volume} {68}},\ \bibinfo
  {pages} {062308} (\bibinfo {year} {2003})}\BibitemShut {NoStop}%
\bibitem [{\citenamefont {Gatteschi}\ and\ \citenamefont
  {Sessoli}(2003)}]{gatteschi_quantum_2003}%
  \BibitemOpen
  \bibfield  {author} {\bibinfo {author} {\bibfnamefont {D.}~\bibnamefont
  {Gatteschi}}\ and\ \bibinfo {author} {\bibfnamefont {R.}~\bibnamefont
  {Sessoli}},\ }\bibfield  {title} {\bibinfo {title} {Quantum tunneling of
  magnetization and related phenomena in molecular materials},\ }\href
  {https://onlinelibrary.wiley.com/doi/abs/10.1002/anie.200390099} {\bibfield
  {journal} {\bibinfo  {journal} {Angew. Chem., Int. Ed.}\ }\textbf {\bibinfo
  {volume} {42}},\ \bibinfo {pages} {268} (\bibinfo {year} {2003})}\BibitemShut
  {NoStop}%
\bibitem [{\citenamefont {Aromí}\ \emph {et~al.}(2012)\citenamefont {Aromí},
  \citenamefont {Aguilà}, \citenamefont {Luis}, \citenamefont {Hill},\ and\
  \citenamefont {Coronado}}]{aromi_design_2012}%
  \BibitemOpen
  \bibfield  {author} {\bibinfo {author} {\bibfnamefont {G.}~\bibnamefont
  {Aromí}}, \bibinfo {author} {\bibfnamefont {D.}~\bibnamefont {Aguilà}},
  \bibinfo {author} {\bibfnamefont {F.}~\bibnamefont {Luis}}, \bibinfo {author}
  {\bibfnamefont {S.}~\bibnamefont {Hill}},\ and\ \bibinfo {author}
  {\bibfnamefont {E.}~\bibnamefont {Coronado}},\ }\bibfield  {title} {\bibinfo
  {title} {Design of magnetic coordination complexes for quantum computing},\
  }\href {https://pubs.rsc.org/en/content/articlelanding/2012/cs/c1cs15115k}
  {\bibfield  {journal} {\bibinfo  {journal} {Chem. Soc. Rev.}\ }\textbf
  {\bibinfo {volume} {41}},\ \bibinfo {pages} {537} (\bibinfo {year}
  {2012})}\BibitemShut {NoStop}%
\bibitem [{\citenamefont {Atzori}\ and\ \citenamefont
  {Sessoli}(2019)}]{atzori_second_2019}%
  \BibitemOpen
  \bibfield  {author} {\bibinfo {author} {\bibfnamefont {M.}~\bibnamefont
  {Atzori}}\ and\ \bibinfo {author} {\bibfnamefont {R.}~\bibnamefont
  {Sessoli}},\ }\bibfield  {title} {\bibinfo {title} {The second quantum
  revolution: Role and challenges of molecular chemistry},\ }\href
  {https://pubs.acs.org/doi/10.1021/jacs.9b00984} {\bibfield  {journal}
  {\bibinfo  {journal} {J. Am. Chem. Soc.}\ }\textbf {\bibinfo {volume}
  {141}},\ \bibinfo {pages} {11339} (\bibinfo {year} {2019})}\BibitemShut
  {NoStop}%
\bibitem [{\citenamefont {Gaita-Ariño}\ \emph {et~al.}(2019)\citenamefont
  {Gaita-Ariño}, \citenamefont {Luis}, \citenamefont {Hill},\ and\
  \citenamefont {Coronado}}]{gaita-arino_molecular_2019}%
  \BibitemOpen
  \bibfield  {author} {\bibinfo {author} {\bibfnamefont {A.}~\bibnamefont
  {Gaita-Ariño}}, \bibinfo {author} {\bibfnamefont {F.}~\bibnamefont {Luis}},
  \bibinfo {author} {\bibfnamefont {S.}~\bibnamefont {Hill}},\ and\ \bibinfo
  {author} {\bibfnamefont {E.}~\bibnamefont {Coronado}},\ }\bibfield  {title}
  {\bibinfo {title} {Molecular spins for quantum computation},\ }\href
  {https://www.nature.com/articles/s41557-019-0232-y} {\bibfield  {journal}
  {\bibinfo  {journal} {Nat. Chem.}\ }\textbf {\bibinfo {volume} {11}},\
  \bibinfo {pages} {301} (\bibinfo {year} {2019})}\BibitemShut {NoStop}%
\bibitem [{\citenamefont {Carretta}\ \emph {et~al.}(2021)\citenamefont
  {Carretta}, \citenamefont {Zueco}, \citenamefont {Chiesa}, \citenamefont
  {G{\'{o}}mez-Le{\'{o}}n},\ and\ \citenamefont
  {Luis}}]{carretta_perspective_2021}%
  \BibitemOpen
  \bibfield  {author} {\bibinfo {author} {\bibfnamefont {S.}~\bibnamefont
  {Carretta}}, \bibinfo {author} {\bibfnamefont {D.}~\bibnamefont {Zueco}},
  \bibinfo {author} {\bibfnamefont {A.}~\bibnamefont {Chiesa}}, \bibinfo
  {author} {\bibfnamefont {{\'{A}}.}~\bibnamefont {G{\'{o}}mez-Le{\'{o}}n}},\
  and\ \bibinfo {author} {\bibfnamefont {F.}~\bibnamefont {Luis}},\ }\bibfield
  {title} {\bibinfo {title} {A perspective on scaling up quantum computation
  with molecular spins},\ }\href
  {https://pubs.aip.org/aip/apl/article/118/24/240501/238981/A-perspective-on-scaling-up-quantum-computation}
  {\bibfield  {journal} {\bibinfo  {journal} {Appl. Phys. Lett.}\ }\textbf
  {\bibinfo {volume} {118}},\ \bibinfo {pages} {240501} (\bibinfo {year}
  {2021})}\BibitemShut {NoStop}%
\bibitem [{\citenamefont {Castro}\ \emph {et~al.}(2022)\citenamefont {Castro},
  \citenamefont {García~Carrizo}, \citenamefont {Roca}, \citenamefont
  {Zueco},\ and\ \citenamefont {Luis}}]{castro_optimal_2022}%
  \BibitemOpen
  \bibfield  {author} {\bibinfo {author} {\bibfnamefont {A.}~\bibnamefont
  {Castro}}, \bibinfo {author} {\bibfnamefont {A.}~\bibnamefont
  {García~Carrizo}}, \bibinfo {author} {\bibfnamefont {S.}~\bibnamefont
  {Roca}}, \bibinfo {author} {\bibfnamefont {D.}~\bibnamefont {Zueco}},\ and\
  \bibinfo {author} {\bibfnamefont {F.}~\bibnamefont {Luis}},\ }\bibfield
  {title} {\bibinfo {title} {Optimal control of molecular spin qudits},\ }\href
  {https://link.aps.org/doi/10.1103/PhysRevApplied.17.064028} {\bibfield
  {journal} {\bibinfo  {journal} {Phys. Rev. Applied}\ }\textbf {\bibinfo
  {volume} {17}},\ \bibinfo {pages} {064028} (\bibinfo {year}
  {2022})}\BibitemShut {NoStop}%
\bibitem [{\citenamefont {Cao}\ \emph {et~al.}(2003)\citenamefont {Cao},
  \citenamefont {Li}, \citenamefont {Petzold},\ and\ \citenamefont
  {Serban}}]{Cao2003}%
  \BibitemOpen
  \bibfield  {author} {\bibinfo {author} {\bibfnamefont {Y.}~\bibnamefont
  {Cao}}, \bibinfo {author} {\bibfnamefont {S.}~\bibnamefont {Li}}, \bibinfo
  {author} {\bibfnamefont {L.}~\bibnamefont {Petzold}},\ and\ \bibinfo {author}
  {\bibfnamefont {R.}~\bibnamefont {Serban}},\ }\bibfield  {title} {\bibinfo
  {title} {{Adjoint Sensitivity Analysis for Differential-Algebraic Equations:
  The Adjoint {DAE} System and Its Numerical Solution}},\ }\href
  {https://doi.org/10.1137/S1064827501380630} {\bibfield  {journal} {\bibinfo
  {journal} {SIAM Journal on Scientific Computing}\ }\textbf {\bibinfo {volume}
  {24}},\ \bibinfo {pages} {1076} (\bibinfo {year} {2003})}\BibitemShut
  {NoStop}%
\bibitem [{\citenamefont {Pontryagin}\ \emph
  {et~al.}(1962{\natexlab{a}})\citenamefont {Pontryagin}, \citenamefont
  {Boltyanskii}, \citenamefont {Gamkrelidze},\ and\ \citenamefont
  {Mishchenko}}]{Pontryagin1962}%
  \BibitemOpen
  \bibfield  {author} {\bibinfo {author} {\bibfnamefont {L.~S.}\ \bibnamefont
  {Pontryagin}}, \bibinfo {author} {\bibfnamefont {V.~G.}\ \bibnamefont
  {Boltyanskii}}, \bibinfo {author} {\bibfnamefont {R.~V.}\ \bibnamefont
  {Gamkrelidze}},\ and\ \bibinfo {author} {\bibfnamefont {E.~F.}\ \bibnamefont
  {Mishchenko}},\ }\href@noop {} {\emph {\bibinfo {title} {{The Mathematical
  Theory of Optimal Processes}}}}\ (\bibinfo  {publisher} {John Wiley \&
  Sons},\ \bibinfo {year} {1962})\BibitemShut {NoStop}%
\bibitem [{\citenamefont {Reich}\ \emph {et~al.}(2014)\citenamefont {Reich},
  \citenamefont {Palao},\ and\ \citenamefont {Koch}}]{Reich2014}%
  \BibitemOpen
  \bibfield  {author} {\bibinfo {author} {\bibfnamefont {D.~M.}\ \bibnamefont
  {Reich}}, \bibinfo {author} {\bibfnamefont {J.~P.}\ \bibnamefont {Palao}},\
  and\ \bibinfo {author} {\bibfnamefont {C.~P.}\ \bibnamefont {Koch}},\
  }\bibfield  {title} {\bibinfo {title} {Optimal control under spectral
  constraints: enforcing multi-photon absorption pathways},\ }\href
  {https://doi.org/10.1080/09500340.2013.844866} {\bibfield  {journal}
  {\bibinfo  {journal} {Journal of Modern Optics}\ }\textbf {\bibinfo {volume}
  {61}},\ \bibinfo {pages} {822} (\bibinfo {year} {2014})}\BibitemShut
  {NoStop}%
\bibitem [{\citenamefont {Lapert}\ \emph {et~al.}(2009)\citenamefont {Lapert},
  \citenamefont {Tehini}, \citenamefont {Turinici},\ and\ \citenamefont
  {Sugny}}]{Lapert2009}%
  \BibitemOpen
  \bibfield  {author} {\bibinfo {author} {\bibfnamefont {M.}~\bibnamefont
  {Lapert}}, \bibinfo {author} {\bibfnamefont {R.}~\bibnamefont {Tehini}},
  \bibinfo {author} {\bibfnamefont {G.}~\bibnamefont {Turinici}},\ and\
  \bibinfo {author} {\bibfnamefont {D.}~\bibnamefont {Sugny}},\ }\bibfield
  {title} {\bibinfo {title} {Monotonically convergent optimal control theory of
  quantum systems with spectral constraints on the control field},\ }\href
  {https://api.semanticscholar.org/CorpusID:13910494} {\bibfield  {journal}
  {\bibinfo  {journal} {Physical Review A}\ }\textbf {\bibinfo {volume} {79}},\
  \bibinfo {pages} {063411} (\bibinfo {year} {2009})}\BibitemShut {NoStop}%
\bibitem [{\citenamefont {Castro}\ \emph {et~al.}(2012)\citenamefont {Castro},
  \citenamefont {Werschnik},\ and\ \citenamefont {Gross}}]{Castro2012}%
  \BibitemOpen
  \bibfield  {author} {\bibinfo {author} {\bibfnamefont {A.}~\bibnamefont
  {Castro}}, \bibinfo {author} {\bibfnamefont {J.}~\bibnamefont {Werschnik}},\
  and\ \bibinfo {author} {\bibfnamefont {E.~K.~U.}\ \bibnamefont {Gross}},\
  }\bibfield  {title} {\bibinfo {title} {{Controlling the Dynamics of
  Many-Electron Systems from First Principles: A Combination of Optimal Control
  and Time-Dependent Density-Functional Theory}},\ }\bibfield  {journal}
  {\bibinfo  {journal} {Physical Review Letters}\ }\textbf {\bibinfo {volume}
  {{109}}},\ \href {https://doi.org/{10.1103/PhysRevLett.109.153603}}
  {{10.1103/PhysRevLett.109.153603}} (\bibinfo {year} {{2012}})\BibitemShut
  {NoStop}%
\bibitem [{\citenamefont {Machnes}\ \emph {et~al.}(2018)\citenamefont
  {Machnes}, \citenamefont {Ass\'emat}, \citenamefont {Tannor},\ and\
  \citenamefont {Wilhelm}}]{Machnes2018}%
  \BibitemOpen
  \bibfield  {author} {\bibinfo {author} {\bibfnamefont {S.}~\bibnamefont
  {Machnes}}, \bibinfo {author} {\bibfnamefont {E.}~\bibnamefont {Ass\'emat}},
  \bibinfo {author} {\bibfnamefont {D.}~\bibnamefont {Tannor}},\ and\ \bibinfo
  {author} {\bibfnamefont {F.~K.}\ \bibnamefont {Wilhelm}},\ }\bibfield
  {title} {\bibinfo {title} {{Tunable, Flexible, and Efficient Optimization of
  Control Pulses for Practical Qubits}},\ }\href
  {https://doi.org/10.1103/PhysRevLett.120.150401} {\bibfield  {journal}
  {\bibinfo  {journal} {Phys. Rev. Lett.}\ }\textbf {\bibinfo {volume} {120}},\
  \bibinfo {pages} {150401} (\bibinfo {year} {2018})}\BibitemShut {NoStop}%
\bibitem [{\citenamefont {Lucarelli}(2018)}]{Lucarelli2018}%
  \BibitemOpen
  \bibfield  {author} {\bibinfo {author} {\bibfnamefont {D.}~\bibnamefont
  {Lucarelli}},\ }\bibfield  {title} {\bibinfo {title} {{Quantum Optimal
  Control via Gradient Ascent in Function Space and the Time-Bandwidth Quantum
  Speed Limit}},\ }\href {https://doi.org/10.1103/PhysRevA.97.062346}
  {\bibfield  {journal} {\bibinfo  {journal} {Phys. Rev. A}\ }\textbf {\bibinfo
  {volume} {97}},\ \bibinfo {pages} {062346} (\bibinfo {year}
  {2018})}\BibitemShut {NoStop}%
\bibitem [{\citenamefont {S\o{}rensen}\ \emph {et~al.}(2018)\citenamefont
  {S\o{}rensen}, \citenamefont {Aranburu}, \citenamefont {Heinzel},\ and\
  \citenamefont {Sherson}}]{Sorensen2018}%
  \BibitemOpen
  \bibfield  {author} {\bibinfo {author} {\bibfnamefont {J.~J. W.~H.}\
  \bibnamefont {S\o{}rensen}}, \bibinfo {author} {\bibfnamefont {M.~O.}\
  \bibnamefont {Aranburu}}, \bibinfo {author} {\bibfnamefont {T.}~\bibnamefont
  {Heinzel}},\ and\ \bibinfo {author} {\bibfnamefont {J.~F.}\ \bibnamefont
  {Sherson}},\ }\bibfield  {title} {\bibinfo {title} {{Quantum Optimal Control
  in a Chopped Basis: Applications in Control of Bose-Einstein Condensates}},\
  }\href {https://doi.org/10.1103/PhysRevA.98.022119} {\bibfield  {journal}
  {\bibinfo  {journal} {Phys. Rev. A}\ }\textbf {\bibinfo {volume} {98}},\
  \bibinfo {pages} {022119} (\bibinfo {year} {2018})}\BibitemShut {NoStop}%
\bibitem [{\citenamefont {Jenkins}\ \emph {et~al.}(2017)\citenamefont
  {Jenkins}, \citenamefont {Duan}, \citenamefont {Diosdado}, \citenamefont
  {García-Ripoll}, \citenamefont {Gaita-Ariño}, \citenamefont
  {Giménez-Saiz}, \citenamefont {Alonso}, \citenamefont {Coronado},\ and\
  \citenamefont {Luis}}]{jenkins_coherent_2017}%
  \BibitemOpen
  \bibfield  {author} {\bibinfo {author} {\bibfnamefont {M.~D.}\ \bibnamefont
  {Jenkins}}, \bibinfo {author} {\bibfnamefont {Y.}~\bibnamefont {Duan}},
  \bibinfo {author} {\bibfnamefont {B.}~\bibnamefont {Diosdado}}, \bibinfo
  {author} {\bibfnamefont {J.~J.}\ \bibnamefont {García-Ripoll}}, \bibinfo
  {author} {\bibfnamefont {A.}~\bibnamefont {Gaita-Ariño}}, \bibinfo {author}
  {\bibfnamefont {C.}~\bibnamefont {Giménez-Saiz}}, \bibinfo {author}
  {\bibfnamefont {P.~J.}\ \bibnamefont {Alonso}}, \bibinfo {author}
  {\bibfnamefont {E.}~\bibnamefont {Coronado}},\ and\ \bibinfo {author}
  {\bibfnamefont {F.}~\bibnamefont {Luis}},\ }\bibfield  {title} {\bibinfo
  {title} {Coherent manipulation of three-qubit states in a molecular
  single-ion magnet},\ }\href
  {https://journals.aps.org/prb/abstract/10.1103/PhysRevB.95.064423} {\bibfield
   {journal} {\bibinfo  {journal} {Phys. Rev. B}\ }\textbf {\bibinfo {volume}
  {95}},\ \bibinfo {pages} {064423} (\bibinfo {year} {2017})}\BibitemShut
  {NoStop}%
\bibitem [{\citenamefont {Martínez-Pérez}\ \emph {et~al.}(2012)\citenamefont
  {Martínez-Pérez}, \citenamefont {Cardona-Serra}, \citenamefont {Schlegel},
  \citenamefont {Moro}, \citenamefont {Alonso}, \citenamefont {Prima-García},
  \citenamefont {Clemente-Juan}, \citenamefont {Evangelisti}, \citenamefont
  {Gaita-Ariño}, \citenamefont {Sesé}, \citenamefont {van Slageren},
  \citenamefont {Coronado},\ and\ \citenamefont
  {Luis}}]{martinez-perez_gd-based_2012}%
  \BibitemOpen
  \bibfield  {author} {\bibinfo {author} {\bibfnamefont {M.~J.}\ \bibnamefont
  {Martínez-Pérez}}, \bibinfo {author} {\bibfnamefont {S.}~\bibnamefont
  {Cardona-Serra}}, \bibinfo {author} {\bibfnamefont {C.}~\bibnamefont
  {Schlegel}}, \bibinfo {author} {\bibfnamefont {F.}~\bibnamefont {Moro}},
  \bibinfo {author} {\bibfnamefont {P.~J.}\ \bibnamefont {Alonso}}, \bibinfo
  {author} {\bibfnamefont {H.}~\bibnamefont {Prima-García}}, \bibinfo {author}
  {\bibfnamefont {J.~M.}\ \bibnamefont {Clemente-Juan}}, \bibinfo {author}
  {\bibfnamefont {M.}~\bibnamefont {Evangelisti}}, \bibinfo {author}
  {\bibfnamefont {A.}~\bibnamefont {Gaita-Ariño}}, \bibinfo {author}
  {\bibfnamefont {J.}~\bibnamefont {Sesé}}, \bibinfo {author} {\bibfnamefont
  {J.}~\bibnamefont {van Slageren}}, \bibinfo {author} {\bibfnamefont
  {E.}~\bibnamefont {Coronado}},\ and\ \bibinfo {author} {\bibfnamefont
  {F.}~\bibnamefont {Luis}},\ }\bibfield  {title} {\bibinfo {title} {Gd-based
  single-ion magnets with tunable magnetic anisotropy: Molecular design of spin
  qubits},\ }\href {https://link.aps.org/doi/10.1103/PhysRevLett.108.247213}
  {\bibfield  {journal} {\bibinfo  {journal} {Phys. Rev. Lett.}\ }\textbf
  {\bibinfo {volume} {108}},\ \bibinfo {pages} {247213} (\bibinfo {year}
  {2012})}\BibitemShut {NoStop}%
\bibitem [{\citenamefont {White}(2007)}]{White2007}%
  \BibitemOpen
  \bibfield  {author} {\bibinfo {author} {\bibfnamefont {R.~M.}\ \bibnamefont
  {White}},\ }\href {https://www.springer.com/gp/book/9783540651161} {\emph
  {\bibinfo {title} {Quantum Theory of Magnetism : Magnetic Properties of
  Materials}}}\ (\bibinfo  {publisher} {Springer Berlin Heidelberg},\ \bibinfo
  {year} {2007})\BibitemShut {NoStop}%
\bibitem [{\citenamefont {Lindblad}(1976)}]{lindblad_generators_1976}%
  \BibitemOpen
  \bibfield  {author} {\bibinfo {author} {\bibfnamefont {G.}~\bibnamefont
  {Lindblad}},\ }\bibfield  {title} {\bibinfo {title} {On the generators of
  quantum dynamical semigroups},\ }\href {https://doi.org/10.1007/BF01608499}
  {\bibfield  {journal} {\bibinfo  {journal} {Commun. Math. Phys.}\ }\textbf
  {\bibinfo {volume} {48}},\ \bibinfo {pages} {119} (\bibinfo {year}
  {1976})}\BibitemShut {NoStop}%
\bibitem [{\citenamefont {Gorini}\ \emph {et~al.}(1976)\citenamefont {Gorini},
  \citenamefont {Kossakowski},\ and\ \citenamefont
  {Sudarshan}}]{gorini_completely_1976}%
  \BibitemOpen
  \bibfield  {author} {\bibinfo {author} {\bibfnamefont {V.}~\bibnamefont
  {Gorini}}, \bibinfo {author} {\bibfnamefont {A.}~\bibnamefont
  {Kossakowski}},\ and\ \bibinfo {author} {\bibfnamefont {E.~C.~G.}\
  \bibnamefont {Sudarshan}},\ }\bibfield  {title} {\bibinfo {title} {Completely
  positive dynamical semigroups of n‐level systems},\ }\href
  {https://pubs.aip.org/aip/jmp/article/17/5/821/225427/Completely-positive-dynamical-semigroups-of-N}
  {\bibfield  {journal} {\bibinfo  {journal} {J. Math. Phys.}\ }\textbf
  {\bibinfo {volume} {17}},\ \bibinfo {pages} {821} (\bibinfo {year}
  {1976})}\BibitemShut {NoStop}%
\bibitem [{\citenamefont {Chiesa}\ \emph {et~al.}(2022)\citenamefont {Chiesa},
  \citenamefont {Petiziol}, \citenamefont {Chizzini}, \citenamefont {Santini},\
  and\ \citenamefont {Carretta}}]{chiesa_theoretical_2022}%
  \BibitemOpen
  \bibfield  {author} {\bibinfo {author} {\bibfnamefont {A.}~\bibnamefont
  {Chiesa}}, \bibinfo {author} {\bibfnamefont {F.}~\bibnamefont {Petiziol}},
  \bibinfo {author} {\bibfnamefont {M.}~\bibnamefont {Chizzini}}, \bibinfo
  {author} {\bibfnamefont {P.}~\bibnamefont {Santini}},\ and\ \bibinfo {author}
  {\bibfnamefont {S.}~\bibnamefont {Carretta}},\ }\bibfield  {title} {\bibinfo
  {title} {Theoretical {Design} of {Optimal} {Molecular} {Qudits} for {Quantum}
  {Error} {Correction}},\ }\href {https://doi.org/10.1021/acs.jpclett.2c01602}
  {\bibfield  {journal} {\bibinfo  {journal} {J. Phys. Chem. Lett.}\ }\textbf
  {\bibinfo {volume} {13}},\ \bibinfo {pages} {6468} (\bibinfo {year}
  {2022})}\BibitemShut {NoStop}%
\bibitem [{\citenamefont {Chiesa}\ \emph {et~al.}(2024)\citenamefont {Chiesa},
  \citenamefont {Santini}, \citenamefont {Garlatti}, \citenamefont {Luis},\
  and\ \citenamefont {Carretta}}]{chiesa_molecular_2024}%
  \BibitemOpen
  \bibfield  {author} {\bibinfo {author} {\bibfnamefont {A.}~\bibnamefont
  {Chiesa}}, \bibinfo {author} {\bibfnamefont {P.}~\bibnamefont {Santini}},
  \bibinfo {author} {\bibfnamefont {E.}~\bibnamefont {Garlatti}}, \bibinfo
  {author} {\bibfnamefont {F.}~\bibnamefont {Luis}},\ and\ \bibinfo {author}
  {\bibfnamefont {S.}~\bibnamefont {Carretta}},\ }\bibfield  {title} {\bibinfo
  {title} {Molecular nanomagnets: a viable path toward quantum information
  processing?},\ }\href {https://doi.org/10.1088/1361-6633/ad1f81} {\bibfield
  {journal} {\bibinfo  {journal} {Rep. Prog. Phys.}\ }\textbf {\bibinfo
  {volume} {87}},\ \bibinfo {pages} {034501} (\bibinfo {year}
  {2024})}\BibitemShut {NoStop}%
\bibitem [{\citenamefont {Janković}\ \emph {et~al.}(2024)\citenamefont
  {Janković}, \citenamefont {Hartmann}, \citenamefont {Ruben},\ and\
  \citenamefont {Hervieux}}]{jankovic_noisy_2024}%
  \BibitemOpen
  \bibfield  {author} {\bibinfo {author} {\bibfnamefont {D.}~\bibnamefont
  {Janković}}, \bibinfo {author} {\bibfnamefont {J.-G.}\ \bibnamefont
  {Hartmann}}, \bibinfo {author} {\bibfnamefont {M.}~\bibnamefont {Ruben}},\
  and\ \bibinfo {author} {\bibfnamefont {P.-A.}\ \bibnamefont {Hervieux}},\
  }\bibfield  {title} {\bibinfo {title} {Noisy qudit vs multiple qubits:
  conditions on gate efficiency for enhancing fidelity},\ }\href
  {https://doi.org/10.1038/s41534-024-00829-6} {\bibfield  {journal} {\bibinfo
  {journal} {npj Quantum Inf.}\ }\textbf {\bibinfo {volume} {10}},\ \bibinfo
  {pages} {1} (\bibinfo {year} {2024})}\BibitemShut {NoStop}%
\bibitem [{\citenamefont {Petiziol}\ \emph {et~al.}(2021)\citenamefont
  {Petiziol}, \citenamefont {Chiesa}, \citenamefont {Wimberger}, \citenamefont
  {Santini},\ and\ \citenamefont {Carretta}}]{petiziol_counteracting_2021}%
  \BibitemOpen
  \bibfield  {author} {\bibinfo {author} {\bibfnamefont {F.}~\bibnamefont
  {Petiziol}}, \bibinfo {author} {\bibfnamefont {A.}~\bibnamefont {Chiesa}},
  \bibinfo {author} {\bibfnamefont {S.}~\bibnamefont {Wimberger}}, \bibinfo
  {author} {\bibfnamefont {P.}~\bibnamefont {Santini}},\ and\ \bibinfo {author}
  {\bibfnamefont {S.}~\bibnamefont {Carretta}},\ }\bibfield  {title} {\bibinfo
  {title} {Counteracting dephasing in molecular nanomagnets by optimized qudit
  encodings},\ }\href {http://dx.doi.org/10.1038/s41534-021-00466-3} {\bibfield
   {journal} {\bibinfo  {journal} {npj Quantum Inf.}\ }\textbf {\bibinfo
  {volume} {7}} (\bibinfo {year} {2021})}\BibitemShut {NoStop}%
\bibitem [{Note1()}]{Note1}%
  \BibitemOpen
  \bibinfo {note} {In this work, we will assume that the purpose is to realize
  a given gate in the interaction representation.}\BibitemShut {Stop}%
\bibitem [{Note2()}]{Note2}%
  \BibitemOpen
  \bibinfo {note} {An implicit assumption in Eq.~(\ref {eq:lambdaT}) (and in
  the discussion around it) is the possibility of discontinouisly switching
  from one monochromatic pulse to another. In practice, these pulses require
  some switch-on and off phases, that make the real implementations slower.
  Expressions (\ref {eq:lambdaT1}) and (\ref {eq:lambdaT}) can thus be
  considered lower bounds for the durations}\BibitemShut {NoStop}%
\bibitem [{\citenamefont {Koch}(2016)}]{Koch_2016}%
  \BibitemOpen
  \bibfield  {author} {\bibinfo {author} {\bibfnamefont {C.~P.}\ \bibnamefont
  {Koch}},\ }\bibfield  {title} {\bibinfo {title} {Controlling open quantum
  systems: tools, achievements, and limitations},\ }\href
  {https://doi.org/10.1088/0953-8984/28/21/213001} {\bibfield  {journal}
  {\bibinfo  {journal} {Journal of Physics: Condensed Matter}\ }\textbf
  {\bibinfo {volume} {28}},\ \bibinfo {pages} {213001} (\bibinfo {year}
  {2016})}\BibitemShut {NoStop}%
\bibitem [{\citenamefont {Schulte-Herbrüggen}\ \emph
  {et~al.}(2011)\citenamefont {Schulte-Herbrüggen}, \citenamefont {Spörl},
  \citenamefont {Khaneja},\ and\ \citenamefont
  {Glaser}}]{Schulte-Herbruggen_2011}%
  \BibitemOpen
  \bibfield  {author} {\bibinfo {author} {\bibfnamefont {T.}~\bibnamefont
  {Schulte-Herbrüggen}}, \bibinfo {author} {\bibfnamefont {A.}~\bibnamefont
  {Spörl}}, \bibinfo {author} {\bibfnamefont {N.}~\bibnamefont {Khaneja}},\
  and\ \bibinfo {author} {\bibfnamefont {S.~J.}\ \bibnamefont {Glaser}},\
  }\bibfield  {title} {\bibinfo {title} {Optimal control for generating quantum
  gates in open dissipative systems},\ }\href
  {https://doi.org/10.1088/0953-4075/44/15/154013} {\bibfield  {journal}
  {\bibinfo  {journal} {Journal of Physics B: Atomic, Molecular and Optical
  Physics}\ }\textbf {\bibinfo {volume} {44}},\ \bibinfo {pages} {154013}
  (\bibinfo {year} {2011})}\BibitemShut {NoStop}%
\bibitem [{\citenamefont {Goerz}\ \emph {et~al.}(2014)\citenamefont {Goerz},
  \citenamefont {Reich},\ and\ \citenamefont {Koch}}]{goerz_optimal_2014}%
  \BibitemOpen
  \bibfield  {author} {\bibinfo {author} {\bibfnamefont {M.~H.}\ \bibnamefont
  {Goerz}}, \bibinfo {author} {\bibfnamefont {D.~M.}\ \bibnamefont {Reich}},\
  and\ \bibinfo {author} {\bibfnamefont {C.~P.}\ \bibnamefont {Koch}},\
  }\bibfield  {title} {\bibinfo {title} {Optimal control theory for a unitary
  operation under dissipative evolution},\ }\href
  {https://dx.doi.org/10.1088/1367-2630/16/5/055012} {\bibfield  {journal}
  {\bibinfo  {journal} {New J. Phys.}\ }\textbf {\bibinfo {volume} {16}},\
  \bibinfo {pages} {055012} (\bibinfo {year} {2014})}\BibitemShut {NoStop}%
\bibitem [{\citenamefont {Pontryagin}\ \emph
  {et~al.}(1962{\natexlab{b}})\citenamefont {Pontryagin}, \citenamefont
  {Boltyanskii}, \citenamefont {Gamkrelidze},\ and\ \citenamefont
  {Mishechenko}}]{pontryagin_mathematical_1962}%
  \BibitemOpen
  \bibfield  {author} {\bibinfo {author} {\bibfnamefont {L.~S.}\ \bibnamefont
  {Pontryagin}}, \bibinfo {author} {\bibfnamefont {V.~G.}\ \bibnamefont
  {Boltyanskii}}, \bibinfo {author} {\bibfnamefont {R.~V.}\ \bibnamefont
  {Gamkrelidze}},\ and\ \bibinfo {author} {\bibfnamefont {E.~F.}\ \bibnamefont
  {Mishechenko}},\ }\href@noop {} {\emph {\bibinfo {title} {The Mathematical
  Theory of Optimal Processes}}}\ (\bibinfo  {publisher} {John Wiley \& Sons},\
  \bibinfo {year} {1962})\BibitemShut {NoStop}%
\bibitem [{\citenamefont {Castro}(2024)}]{castro_qocttools_2024}%
  \BibitemOpen
  \bibfield  {author} {\bibinfo {author} {\bibfnamefont {A.}~\bibnamefont
  {Castro}},\ }\bibfield  {title} {\bibinfo {title} {qocttools: A program for
  quantum optimal control calculations},\ }\href
  {https://www.sciencedirect.com/science/article/pii/S0010465523003284}
  {\bibfield  {journal} {\bibinfo  {journal} {Comput. Phys. Commun.}\ }\textbf
  {\bibinfo {volume} {295}},\ \bibinfo {pages} {108983} (\bibinfo {year}
  {2024})}\BibitemShut {NoStop}%
\bibitem [{\citenamefont {Akima}(1970)}]{Akima1970}%
  \BibitemOpen
  \bibfield  {author} {\bibinfo {author} {\bibfnamefont {H.}~\bibnamefont
  {Akima}},\ }\bibfield  {title} {\bibinfo {title} {A new method of
  interpolation and smooth curve fitting based on local procedures},\ }\href
  {https://doi.org/10.1145/321607.321609} {\bibfield  {journal} {\bibinfo
  {journal} {J. ACM}\ }\textbf {\bibinfo {volume} {17}},\ \bibinfo {pages}
  {589–602} (\bibinfo {year} {1970})}\BibitemShut {NoStop}%
\end{thebibliography}%

\end{document}